\documentclass[twocolumn]{aastex631}
\received{2022 November 26}
\accepted{2022 December 28}

\shorttitle{OJ 287}
\shortauthors{S. Komossa et al.}
\graphicspath{{./}{figures/}}
\begin{document}

\title{MOMO VI: Multifrequency radio variability of the blazar OJ 287 from 2015--2022, absence of predicted 2021 
precursor-flare activity, and a new binary interpretation of the 2016/2017 outburst}

%% LaTeX will automatically break titles if they run longer than
%% one line. However, you may use \\ to force a line break if
%% you desire. In v6.31 you can include a footnote in the title.

%% The \author command is the same as before except it now takes an optional
%% argument which is the 16 digit ORCID. The syntax is:
%% \author[xxxx-xxxx-xxxx-xxxx]{Author Name}
%%
%% This will hyperlink the author name to the author's ORCID page. 
%%
%\correspondingauthor{August Muench}
%\email{greg.schwarz@aas.org, gus.muench@aas.org}

%\author[0000-0002-0786-7307]{G. Mueller}
%\affiliation{American Astronomical Society \\
%1667 K Street NW, Suite 800 \\
%Washington, DC 20006, USA}

\author{S. Komossa}
\affiliation{Max-Planck-Institut f{\"u}r Radioastronomie, 
Auf dem H{\"u}gel 69, 53121 Bonn, Germany}

\author{A. Kraus}
\affiliation{Max-Planck-Institut f{\"u}r Radioastronomie, Auf dem H{\"u}gel 69, 53121 Bonn, Germany}

\author{D. Grupe}
\affiliation{Department of Physics, Geology, and Engineering Technology, Northern Kentucky University, 1 Nunn Dr, Highland Heights, KY 41099, USA}

\author{A.G. Gonzalez}
\affiliation{Department of Astronomy and Physics, Saint Mary’s University, 923 Robie Street, Halifax, NS, B3H 3C3, Canada}

%%% \vert{} 
\author{M.A. Gurwell}
\affiliation{Center for Astrophysics, Harvard \& Smithsonian, Cambridge, MA 02138, USA}

\author{L.C. Gallo}
\affiliation{Department of Astronomy and Physics, Saint Mary’s University, 923 Robie Street, Halifax, NS, B3H 3C3, Canada}

\author{F.K. Liu}
\affiliation{Department of Astronomy, School of Physics, Peking University, Beijing 100871, People’s Republic of China}
\affiliation{Kavli Institute for Astronomy and Astrophysics, Peking University, Beijing 100871, People’s Republic of China}

\author{I. Myserlis} 
\affiliation{Institut de Radioastronomie Millim\'etrique, Avenida Divina Pastora 7, Local 20, 18012, Granada, Spain}
\affiliation{Max-Planck-Institut f{\"u}r Radioastronomie, Auf dem H{\"u}gel 69, 53121 Bonn, Germany}

\author{T.P. Krichbaum}
\affiliation{Max-Planck-Institut f{\"u}r Radioastronomie, Auf dem H{\"u}gel 69, 53121 Bonn, Germany}

\author{S. Laine}
\affiliation{IPAC, Mail Code 314-6, Caltech, 1200 E. California Blvd., Pasadena, CA 91125, USA} 

\author{U. Bach}
\affiliation{Max-Planck-Institut f{\"u}r Radioastronomie, Auf dem H{\"u}gel 69, 53121 Bonn, Germany}

\author{J.L. G\'omez}
\affiliation{Instituto de Astrofísica de Andalucía-CSIC, Glorieta de la Astronomía s/n, E-18008 Granada, Spain} 

\author{M.L. Parker}
\affiliation{Institute of Astronomy, University of Cambridge, Madingley Road, Cambridge CB3 0HA, UK}

\author{S. Yao}
\affiliation{Max-Planck-Institut f{\"u}r Radioastronomie, Auf dem H{\"u}gel 69, 53121 Bonn, Germany}

\author{M. Berton}
\affiliation{European Southern Observatory (ESO), Alonso de Córdova 3107, Casilla 19, Santiago 19001, Chile}

%\author{et al.}
%\affiliation{...}

%% AASTeX 6.31 has the new \collaboration and \nocollaboration commands to
%% provide the collaboration status of a group of authors. These commands 
%% can be used either before or after the list of corresponding authors. The
%% argument for \collaboration is the collaboration identifier. Authors are
%% encouraged to surround collaboration identifiers with ()s. The 
%% \nocollaboration command takes no argument and exists to indicate that
%% the nearby authors are not part of surrounding collaborations.

%% Mark off the abstract in the ``abstract'' environment. 
%% abstract limited to 250 words. 
\begin{abstract}
Based on our dedicated Swift monitoring program, MOMO, OJ 287 is one of the best-monitored blazars in the X-ray--UV--optical regime.
Here, we report results from our accompanying, dense, multi-frequency (1.4--44 GHz) radio monitoring of OJ 287 between 2015 and 2022 
covering a broad range of activity states. Fermi $\gamma$-ray observations are added. 
We characterize the radio flux and spectral variability in detail, including DCF and other variability analyses, and discuss its connection with the multiwavelength emission. 
Deep fades of radio and optical--UV fluxes are found to occur every 1--2 years.  
Further, it is shown that a precursor flare
of thermal bremsstrahlung 
predicted by
one of the binary supermassive black hole (SMBH) models of OJ 287 was absent. 
We then focus on the nature of the extraordinary, nonthermal 2016/2017 outburst that we initially discovered 
with Swift.  We interpret it 
as the latest of the famous optical double-peaked outbursts of OJ 287, favoring binary scenarios that do not require a highly precessing secondary SMBH. 
\end{abstract}


\section{Introduction}

Active galactic nuclei (AGN) are powered by accretion of matter onto supermassive black holes (SMBHs) at their centers \citep{Lynden-Bell1969}. 
A fraction of AGN is radio-loud and harbors powerful, long-lived jets of relativistic particles that are launched in the immediate vicinity of their central SMBHs.  The SMBH -- accretion disk -- jet interface represents one of the most extreme astrophysical environments where magnetic fields, high gas density, and (special and general) relativistic astrophysics all play a crucial role in shaping the multiwavelength electromagnetic emission and structural properties of these systems \citep[see][for a recent review]{Blandford2019}.

In blazars, the jet is oriented close to the observer's line of sight, and its emission is enhanced by relativistic beaming. 
The spectral energy distribution (SED) of blazars is characterized by different emission processes at different wavelengths.     
SEDs show two broad emission humps \citep[][]{Abdo2010, Ghisellini2017}:
one at low energies peaking between the submillimeter and EUV, sometimes extending into the soft X-ray band, and explained as synchrotron radiation of a population of accelerating jet electrons, and a second maximum in the hard X-ray/$\gamma$-ray regime, usually explained as inverse Compton (IC) radiation from photons that scatter off the jet electrons. 
These photons are located either inside the jet  (synchrotron-self-Compton radiation; SSC) or they are emitted by an external medium such as the broad-line region (BLR) or torus (external Comptonization; EC). Additionally, or alternatively, hadronic processes (ultra-relativistic protons) may contribute at high energies \citep[e.g.,][]{Boettcher2019}. 

OJ 287 is a key representative of the class of blazars. It is relatively nearby (redshift $z=0.306$), very bright across the electromagnetic spectrum, highly variable, and highly polarized \citep[e.g.,][and references therein]{Goddi2021, Komossa2021c, Valtonen2021, Abdollahi2022, Komossa2022a}.  
It exhibits exceptional optical flares that reached as bright as 12th magnitude \citep{Kinman1971}.
% and always show a characteristic double-peak structure. 
OJ 287 has been detected in the $\gamma$-ray and very-high energy (VHE) regime \citep{Abdo2009, OBrien2017}. 
Due to a very dense and dedicated multiwavelength Swift (X-ray--UV--optical) monitoring campaign that we set up $>6$ yrs ago in the course of the project MOMO \citep[Multiwavelength Observations and Modelling of OJ 287;][]{Komossa2021d}, OJ 287 has one of the best-covered long-term X-ray--UV--optical light curves and {{\em simultaneous}} SEDs of any blazar \citep{Komossa2021c}, along with the dense multifrequency radio monitoring presented here.   

On the basis of the faintness of its optical emission lines from the broad-line region, only occasionally detected \citep{SitkoJunkkarinen1985, Nilsson2010}, OJ 287 is classified as a BL Lac object.
OJ 287 caught particular attention following its bright and long-lasting optical outbursts in the 1970s and 80s. 
After strong evidence was presented that these outbursts repeat every $\sim$11--12 yr \citep{Sillanpaa1988} 
and are double-peaked with a peak separation of order 1 yr \citep{Sillanpaa1996}, several binary black hole scenarios were suggested \citep[see][for a recent review]{Komossa2021b}, involving:   
(1) a mildly precessing 
secondary SMBH that is nearly in the accretion-disk plane and is tidally perturbing the primary SMBH's accretion disk near periastron \citep{Sillanpaa1988}; or (2) a highly inclined, highly precessing secondary SMBH impacting the primary's accretion disk twice during each orbit \citep[precessing binary (PB) model hereafter;][]{Lehto1996, Valtonen2021}; or (3) a precessing accretion disk \citep{Katz1997}; or (4) Doppler boosting of two jetted SMBHs orbiting each other \citep{Villata1998};  or (5) a secondary SMBH that impacts the primary's accretion disk only once during its orbit with subsequent tidal perturbations feeding the inner jet \citep{Valtaoja2000}. While the PB model required a very massive primary SMBH of order 10$^{10}$ M$_{\odot}$, \citet{Liu2002} explored
a variant of the 
\citet{Valtaoja2000} model that only required a primary's mass of order 10$^{8}$ M$_{\odot}$. 

The pattern of variability of OJ 287 is not strictly periodic, but comes with deviations from strict periodicity of order 1 yr \citep{Valtaoja2000}. One of the largest deviations happened in 2005--2006 when the brightest outburst occurred in 2005 October--November \citep{Villforth2010} while expected only in 2006.

The best explored binary scenario is the PB model, based on detailed orbital modelling of the system.  
Parameters of that model were adjusted 
each time a supposed `main outburst' was observed at a different epoch than expected, then a new prediction was made, and so on.  
The model claims very high accuracy in the predicted timing of subsequent flares on the order of days -- hours, for events that are separated by more than a decade.

Non-binary models producing semi-periodicity in AGN light curves in general have also been explored \citep[e.g.][]{Rieger2004, Liska2018}, some directly applied to OJ 287: 
\citet{Britzen2018} favored disk precession, leaving open the cause for precession (that could still be a binary), and 
\citet{Villforth2010} presented cautious comments on a binary interpretation and  
speculated about  
a model of magnetic breathing to explain the optical double-peaked outbursts of OJ 287.  

Therefore, in addition to understanding 
disk-jet physics of this bright blazar, the project MOMO was carefully 
designed for rigorously distinguishing between different binary (and non-binary) SMBH scenarios of OJ 287 in general, and testing predictions of the PB model in particular, including distinctly different predictions made in the radio regime (Section 2).  
While most previous results focussed on the UV--optical, X-ray and $\gamma$-ray regime 
\citep[e.g.,][]{Komossa2020, Komossa2021a, Komossa2022a}, 
here we concentrate on the complete 2015--2022 multifrequency radio observations and their interpretation and implications.  

This paper is organized as follows: In Section 2 we describe the key motivation and goals of the MOMO project with emphasis on the questions that will be addressed with the radio multifrequency observations presented here. Sections 3--6 provide a description of the data acquisition and analysis in the radio, optical, UV, X-ray, and $\gamma$-ray bands. In Section 7, the radio flux variability and spectral properties are established and discussed in detail, including DCF analyses, followed by a comparison with the multiwavelength light curves in Section 8. 
Results are discussed in Section 9 with particular focus on timing results, the extraordinary 2016/2017 outburst, and constraints on binary SMBH scenarios. A summary and conclusions are provided in Section 10. 

Timescales and frequencies are given in the observer's frame when reporting measurement results, unless noted otherwise. 
We use a cosmology with 
$H_{\rm 0}$=70 km\,s$^{-1}$\,Mpc$^{-1}$, $\Omega_{\rm M}$=0.3 and $\Omega_{\rm \Lambda}$=0.7 throughout this paper. At the distance of OJ 287, this corresponds to a scale of 4.5 kpc/arcsec \citep{Wright2006}. 

\section{Project MOMO}

\subsection{Overall introduction, goals, and previous results}

In the course of the project MOMO 
\citep[]{Komossa2017, Myserlis2018, Komossa2020, Komossa2021a, Komossa2021c, Komossa2021d, Komossa2022a, Komossa2022b},
we are carrying out dedicated, dense, long-term monitoring of OJ 287 at $>$16 frequencies from radio to X-rays. 
This is the densest monitoring project of OJ 287 so far carried out involving X-rays. MOMO provides spectral and timing information, and simultaneous SEDs, at all activity states of OJ 287.  
In particular, we are using the Effelsberg 100\,m radio telescope and the Neil Gehrels Swift observatory \citep[Swift hereafter;][]{Gehrels2004} since 2015 December. 
Essentially all Swift observations of OJ 287 in recent years were obtained by us in the course of the MOMO program.  
The rich data sets and their interpretation are presented in a sequence of publications and so far include:  
(1) Our detection of two major nonthermal X-ray--UV--optical outbursts with Swift in 2016/17, and 2020
\citep{Komossa2017, Komossa2020}. (2) The detection of variable radio polarization in 2016 \citep{Myserlis2018}. 
(3) The rapid follow-up of the 2020 outburst with Swift, XMM-Newton
and NuSTAR establishing the spectral components up to $\sim$ 70 keV, including an exceptionally strong soft X-ray excess of synchrotron origin and an unexpectedly steep spectrum in the NuSTAR band \citep{Komossa2020}.  (4) XMM-Newton and Swift spectroscopy during the 2018 Event Horizon Telescope (EHT)  campaign \citep{Komossa2021a}. 
(5) A spectral analysis of all XMM-Newton observations during the last two decades \citep{Komossa2021a}, and a spectral and detailed timing analysis of all Swift observations during the last two decades \citep{Komossa2021c}, establishing OJ 287 as one of the most X-ray spectrally variable blazars known, and with strong implications for emission mechanisms.
(6) Effelsberg observations between 2019 and 2021 in a period of high activity \citep{Komossa2022a}, and implying that the radio emission is dominated by the main jet.  
(7) First results on the absence of precursor flare activity \citep{Komossa2022b}, falsifying predictions of one of the binary SMBH model variants.  

\begin{table*}
%\footnotesize
	\centering
	\caption{Receivers used in our Effelsberg 100\,m telescope observations of OJ 287 in the course of the MOMO program since 2015 (Effelsberg program IDs 99-15, 19-16, 12-17, 13-18, 75-19, 65-20, 70-21, and 83-22). 
    Note that some receivers have changed over the years, and at some epochs a larger number of frequencies was observed, partly depending on weather conditions. $\nu_{\rm c}$ is the central frequency at which the flux density was extracted, $\Delta\nu$ the band width, and HPBW the half power beam width. Since the employed central frequency of some bands changes with time, we assign a fixed representative frequency $\nu_{\rm *}$ (as listed in column 5) in the text and in labels of figures. For any analysis and plots, the exact frequencies are used.     
	}
	\label{tab:obs-log-radio}
\begin{tabular}{lccccl}
\hline
Receiver & $\nu_{\rm c}$  & $\Delta \nu$ & HPBW & $\nu_{\rm *}$ & Comment\\
 &  [GHz] &  [MHz] & [arcsec] & \\
\hline
P217mm &  1.40  & 40 & 590 & &  prime focus receiver, used only
occasionally for this project \\
S110mm & 2.595 & 10 & 286 &  2.6 & 2.64 GHz until 2018 May 27 \\
S60mm & 4.85 & 500 & 150 & & \\
S28mm & 10.45 & 300 & 67.5 & & \\
S20mm-old & 14.60 & 2000 & 50 & 14 & used until 2018 March 4; no longer operational \\
S20mm-new & 14.25 & 2500 & 52.9 & 14 &  \\
     & 16.75 & 2500 & 43.7 & & \\
S14mm & 19.25 & 2500 & 40.1 & 19 & 19.20 GHz until 2018 February 10; occasional RFI at 19 GHz \\
    & 21.15 & 2500 & 38.0 & &\\
    & 22.85 & 2500 & 36.8 & &\\
    & 24.75 & 2500 & 33.1 & &\\
S9mm  & 32.00 & 4000 & 24.5 & 36 & used until 2017 October 3\\
S7mm-new  & 36.25 & 2500 & 23.0 & 36 & 35.75 GHz between 2018 April 21 - 2020 May 12  \\
    & 38.75 & 2500 & 21.2 & & 38.25 GHz until 2020 May 12 \\
    & 41.25 & 2500 & 20.7 & & used since 2021 March 24\\
    & 43.75 & 2500 & 19.7 & 43 & 42.75 GHz between 2017 February 01--2018 November 25 \\ 
S7mm-old & 43.00 & 2000 & 20 & 43 & used until 2017 January 3; no longer operational  \\
\hline
\end{tabular}
\end{table*}
%% notes: geringe Bandbreite bei S110 wg RFI

Data we obtain with Swift, the Effelsberg telescope, and at other facilities are analyzed quickly, typically within days, and the community is then rapidly alerted about exceptional flux states (outbursts, deep fades) and spectral states of OJ 287 in a sequence of {\sl{Astronomer's Telegrams}} since 2015 (ATel \#8411, \#9629, \#10043, \#12086, \#13658, \#13702, \#13785, \#14052, \#15145, and \#15764). 
This allows independent data to be taken by the community at frequencies or with instruments not covered by our own program. For instance, the bright 2016/2017 X-ray outburst we detected with Swift \citep[e.g., ATel \#9629, \#10043;][]{Komossa2017, Komossa2020} was used to trigger a VERITAS very-high energy (VHE) observation \citep{OBrien2017}, and the bright 2020 X-ray outburst we detected with Swift (ATel \#13658) triggered an AstroSAT observation \citep{Prince2021, Singh2022} in addition to our own follow-ups with XMM-Newton and NuSTAR \citep{Komossa2020}.      

\subsection{MOMO-radio}

Observations in the radio regime allow us to distinguish between different binary SMBH scenarios in general, and to test predictions of the well-developed PB model in particular, based on distinctly different predictions they make in the radio regime for the timing and properties of the  first and second optical peak of the double-peaked outbursts and for any possible after-flares \citep{Valtaoja2000}.   
For instance, if major optical outbursts 
are thermal in nature, they will not be accompanied by radio (synchrotron)  outbursts{\footnote{Except for possible epochs, where a secondary SMBH may be temporarily accreting and launching a short-lived jet, even though such hypothetical 
events are unlikely to exceed the radio emission from the main jet \citep{Pihajoki2013}}.} If both optical peaks are due to synchrotron emission, we expect two radio flares, with polarization evolution following synchrotron theory.  
These tests require long-term coverage over at least a significant fraction of one binary orbital period. The long-term goal of MOMO is to cover at least two orbital periods. 

Further, independent of the binary hypothesis and of blazar physics directly traced by the Effelsberg observations themselves, we note that the flux and spectral-index measurements from the dense multifrequency radio coverage allow extrapolation to other radio frequencies of interest in deep radio imaging studies of OJ 287 that continue to be carried out with VLBI techniques \citep[e.g.,][]{Lister2021, Weaver2022, Zhao2022}. 

Prior to MOMO, OJ 287 has been the target of other single-dish radio monitoring campaigns, for instance at UMRAO \citep{Hughes1998}, Mets\"ahovi \citep{Hovatta2008}, and Effelsberg \citep{Fuhrmann2016}. These did not include simultaneous UV and X-ray data.

\section{Effelsberg data acquisition and reduction}

\begin{figure*}
\centering
\includegraphics[clip, trim=2.0cm 1.3cm 2.0cm 1.4cm, width=18.0cm]{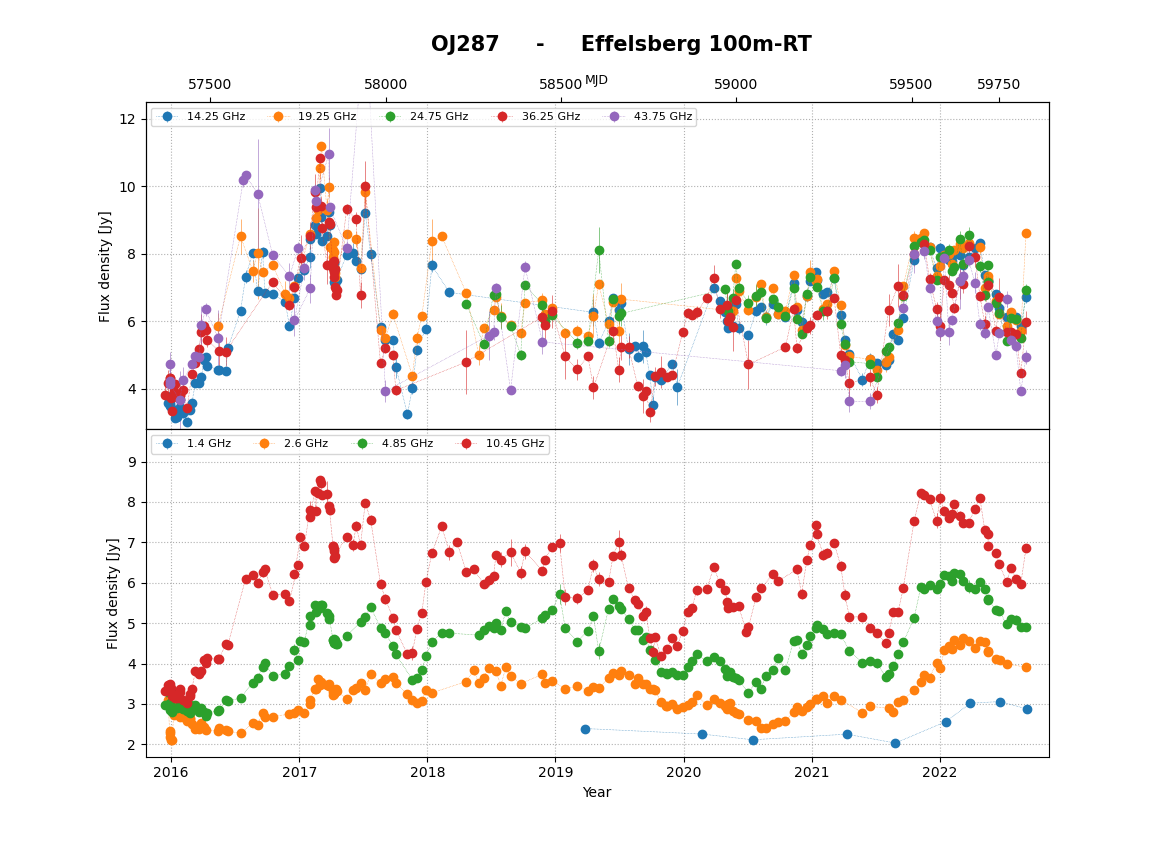}
\caption{Multifrequency radio light curves of OJ 287 between 2015 December and 2022 June obtained with the Effelsberg telescope in the course of the MOMO program. Note that some receivers and/or frequencies have slightly changed in the course of the monitoring. See Tab. \ref{tab:obs-log-radio} for details. }
    \label{fig:radio}
\end{figure*}

We started monitoring OJ 287 in the radio band with the Effelsberg 100\,m telescope \citep{Kraus2003} in December 2015 in the course of the MOMO program. Here we discuss all the data obtained  until June 2022.

OJ 287 is unobservable each year for about 3 months with ground-based optical telescopes, or space-based missions such as Swift, because of its close proximity to the Sun.  Our radio observations have the advantage that OJ 287 can be observed over nearly the whole year. The Sun constraint of the Effelsberg telescope is only 2 degrees, so that sources at larger than 2 degrees separation from the Sun are observable.  

The {\em{average}} cadence of our radio observations is 14 days (at 10.45 GHz). The cadence is higher at selected epochs, and our proposals contained a ToO mode to obtain denser coverage, should some exceptional variability be observed. 
We sometimes experience larger gaps, depending on times of telescope maintenance, availability of receivers, and weather conditions; the latter limiting the cadence for the observations at the highest frequencies (above $\sim$20 GHz).

Data from the first year of observations we already presented in \citet{Myserlis2018}, and data between 2019 to 2021 May were discussed by \citet{Komossa2022a}. These data are re-added here for completeness of discussion, and new analyses are performed on them. 

Observations were carried out between 2.6 and 44 GHz, and occasionally at 1.4 GHz, switching between up to eight receivers. Some of the receivers changed in the course of the monitoring program (see Tab. \ref{tab:obs-log-radio} for details).{\footnote{See \url{https://eff100mwiki.mpifr-bonn.mpg.de/doku.php?id=information_for_astronomers:rx_list} for further technical details on all receivers.}} 
Since the central frequency we employed varied in a few cases over the length of the observing program, we assigned a fixed representative frequency $\nu_{*}$ throughout the text and in some figure labels (for instance, 2.595 and 2.64 GHz are referred to in the text as 2.6 GHz, etc). The exact frequency at any given date can be found in Tab.  \ref{tab:obs-log-radio}.

The cross-scan method \citep{Kraus2003} was used to acquire the radio data.
In the cross-scans, the telescope was moved in two perpendicular directions, azimuth and elevation, with the target source position at the center of the scans. 
For the low frequencies, 2--3 sub-scans were carried out per direction. For the high frequencies,  
it was 12--16 sub-scans per direction.  

The data reduction and analysis was performed in a standard manner as described in, e.g., \citet{Kraus2003}. A detailed description of the analysis procedures will be given by Kraus et al. (in preparation). 
In a first analysis step, the data of every single sub-scan were fit with a Gaussian profile. 
Bad sub-scans were identified and excluded from further analysis. Such bad data sets can be due to high pointing errors, radio frequency interference (RFI), or --- in case of OJ 287 --- disturbances by solar radiation around August 1st of each year due to the source's close proximity to the sun, for instance.  

After correcting for small pointing errors of the telescope, the amplitudes of the individual sub-scans were averaged. In some cases, especially at the highest frequencies and in mediocre
weather conditions, the sub-scans of one direction (azimuth/elevation) were averaged before the Gaussian profile was fit, in order to increase the signal to noise ratio.
Next, corrections for the atmosphere’s opacity were applied as well as for the gain-elevation effect (the change of the sensitivity with elevation). 
Absolute flux calibration was then achieved by comparing the observed antenna temperatures with the expected flux densities of selected calibrators such as 3C 286 \citep{Ott1994}{\footnote{We note in passing that the radio calibrator 3C 286 has recently shown variability in the X-ray and $\gamma$-ray band \citep{YaoKomossa2021}, allowing for the possibility of radio variability of the inner jet, too. However, the bulk of the radio emission of 3C 286 is widely extended and the source is at high redshift, and therefore low-resolution radio observations as the ones carried out here will be unaffected by any possible variability of the inner jet emission. In contrast, in the context of regular monitoring we noticed slight variations 
in  3C 48 and 3C 161, for which we corrected in the course of this project.
At most epochs, more than one calibrator was observed for a cross-check.}}. 
The complete 2015--2022 Effelsberg radio light curves are shown in Fig. \ref{fig:radio}. 

\section{Submillimeter Array (SMA) data acquisition and reduction}

\begin{figure}
\centering
\includegraphics[clip, trim=1.7cm 5.5cm 1.7cm 9.3cm, width=\columnwidth]{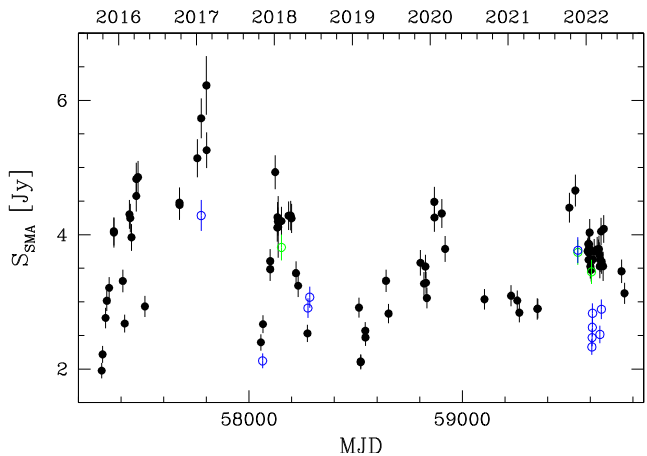}
\caption{SMA radio light curve of OJ 287 between 2015 October and 2022 June (filled circles: 1.3 mm band, open green circles: 1.1 mm band, open blue circles: 870 $\mu$m band). }
    \label{fig:lc-SMA}
\end{figure}

% Standard-Text: 
Flux density data at 1.3 mm, and occasionally at
1.1 mm and 870 $\mu$m, were obtained at the Submillimeter Array (SMA) near the summit of Maunakea (Hawaii). 
OJ 287 is included in an ongoing monitoring program at the SMA to determine the ﬂuxes of compact extragalactic radio sources that can be used as calibrators at mm wavelengths \citep{Gurwell2007}.  Observations of available potential calibrators are from time to time observed for three to five minutes, with the measured source signal strength calibrated against known standards, typically solar system objects (Titan, Uranus, Neptune, or Callisto).  Data from this program are updated regularly and are available at the SMA website{\footnote{http://sma1.sma.hawaii.edu/callist/callist.html}}.  Here, we show the data from 2015 October to 2022 June (Fig. \ref{fig:lc-SMA}). 
The majority of the data in the 1.3 mm band are taken at mean frequencies near 225 GHz, but the exact frequency can vary between 213 GHz and 240 GHz, depending on the science goal of the initial observation. Frequencies in the 1.1mm band of the data reported here range between 270--272 GHz, and in the 870$\mu$m band between 339--351 GHz.  

\section{Swift data acquisition and reduction}

The majority of Swift data we obtained in the course of the MOMO-UO (UV and optical) and MOMO-X (X-rays) program was presented in previous work since 2017, including a detailed discussion of the full 2005--2021 Swift light curve of OJ 287 by \citet{Komossa2021c}. 
Here, we newly present the most recent Swift observations since 2022 mid-February (Tab. \ref{tab:obs-log}). 

The Swift UV--optical telescope \citep[UVOT;][]{Roming2005} and X-ray telescope \citep[XRT;][]{Burrows2005} data reduction is carried out in a standard manner and is described in great detail by \citet{Komossa2021c}. It includes target source and background selection, instrumental corrections, and correction for Galactic extinction of the UVOT fluxes. In the X-ray band, absorption-corrected fluxes are derived from single power-law spectral fits with fixed Galactic absorption. 

\begin{table}
\scriptsize
	\centering
	\caption{Log of Swift observations since 2022 February 19 with OBS-IDs 35905-191 -- 35905-220. $\Delta t$ is the exposure time of each single XRT or UVOT observation. Each UVOT band V:B:U:W1:M2:W2 is observed with a ratio of 1:1:1:2:3:4 of the total exposure time, respectively.}
	\label{tab:obs-log}
	\begin{tabular}{lclc} 
		\hline
		Instrument & Band & Date & $\Delta t$ (ks)\\
		\hline
		XRT & 0.3-10 keV & Feb. -- June 2022 & 1.0--2.2 \\  
		UVOT & V,B,U,W1,M2,W2 & Feb. -- June 2022
		& 1.0--2.2 \\
		\hline
%\begin{tablenotes}
%{Notes: $^{1}$ }.
%\end{tablenotes}
	\end{tabular}
\end{table}

\section{Fermi $\gamma$-ray data}

In the $\gamma$-ray band, Fermi LAT \citep[Large Area Telescope;][]{Atwood2009} data are used. Unlike the Effelsberg and Swift observations that are proposed and analyzed by us, the $\gamma$-ray data were taken from the Fermi archive. They cover the energy regime 0.1--100 GeV. 
Publicly available Fermi LAT data of OJ 287 were retrieved from the Fermi LAT light-curve repository \citep{Kocevski2021}{\footnote{\url{https://fermi.gsfc.nasa.gov/ssc/data/access/lat/LightCurveRepository/}}}. We used weekly averages of the fluxes derived from the spectral model fit of 
a logarithmic parabolic power-law with fixed
parameters (including a fixed photon index of 2.16) and free normalization, as described by \citet{Kocevski2021}. 
This approach is preferred in our case over fits with free index \citep[e.g.][]{Hodgson2017, Kapanadze2018}, because OJ 287 remains in a low $\gamma$-ray state most of the time during the epochs of interest.   

\section{Radio flux and spectral variability}

\subsection{Radio light curves}

The 2015--2022 radio light curves are shown in Fig. \ref{fig:radio}. Several bright radio flares are present. The brightest occurred in 2016--2017 with peak in February 2017 at a flux density of 10.8 Jy at 32 GHz. The radio outburst coincides with the bright and long lasting X-ray--UV--optical outburst of OJ 287, that was detected in the course of our MOMO Swift monitoring \citep{Komossa2017, Komossa2020}. During this time, we see the highest-amplitude of radio variability in recent years. The flux density rises from a deep low-state in the radio emission in 2016 January
by a factor of 3.2 (at 32 GHz) to the observed peak in February 2017. 
A similar rise is seen in the SMA band with an amplitude of a factor of 2.3 between January and the highest measured value in February. 
Due to the lower cadence with no coverage of the flare beyond 2017 February 16, the true peak of the flare cannot be measured at 225 GHz. 
 
Several more radio flares are detected. The second-brightest is observed in 2021 November to 2022 May \citep[see also][]{Komossa2022b}. The 36 GHz flux density rises by a factor 2.2 from minimum to peak. As the flare evolves, the 36 and 44 GHz light curves show a fast decline and renewed rise  to a second maximum in 2022 February. Sharp variability at that epoch is also seen in the SMA data. The radio flux starts to decline in 2022 May. 

Every $\sim$2 years, deep radio minima are observed. 
The lowest states are found in 
late 2015, late 2017 \citep[coincident with the optical--UV deep fade;][]{Komossa2021c}, 
late 2019, and mid 2021.

The most rapid flaring is seen at high frequencies. Flares are often smoothed into broader structures at low frequencies due to opacity effects; see, for instance, the rapid dipping and recovery at high frequencies during the 2022 flaring that is not observed at lower frequencies.  In order to quantify the variability in each frequency band, a fractional variability amplitude analysis was carried out, which we discuss next. 

\subsection{Fractional variability amplitude}

The fractional rms variability amplitude $F_{\rm var}$ \citep{Edelson2002, Vaughan2003} of the radio flux density was calculated separately for each radio frequency at annual time bins (Tab. \ref{tab:Fvar}) and as a long-term average. 

$F_{\rm var}$ is given by:
\begin{equation}
	F_{\rm var} = \sqrt{S^2 - \overline{\sigma^{2}_{\rm err}} \over \overline{x}^2},
\end{equation}
where $S^{2}$ is the variance of the light curve, $\overline{\sigma^{2}_{\rm err}}$ is the mean square of the measurement errors, and $\overline{x}$ is the mean flux \citep{Vaughan2003}. 

The error of $F_{\rm var}$ was calculated according 
to \citet{Edelson2002},  
as: 
\begin{equation}
	\sigma_{F_{\rm var}} = \frac{1}{F_{\rm var}} \sqrt{\frac{1}{2N}} \frac{S^2}{\overline{x}^2},
\end{equation}
where $N$ is the number of data points used in the computation of $F_{\rm var}$.
 
\begin{figure}
\centering
\includegraphics[clip, width=\columnwidth]{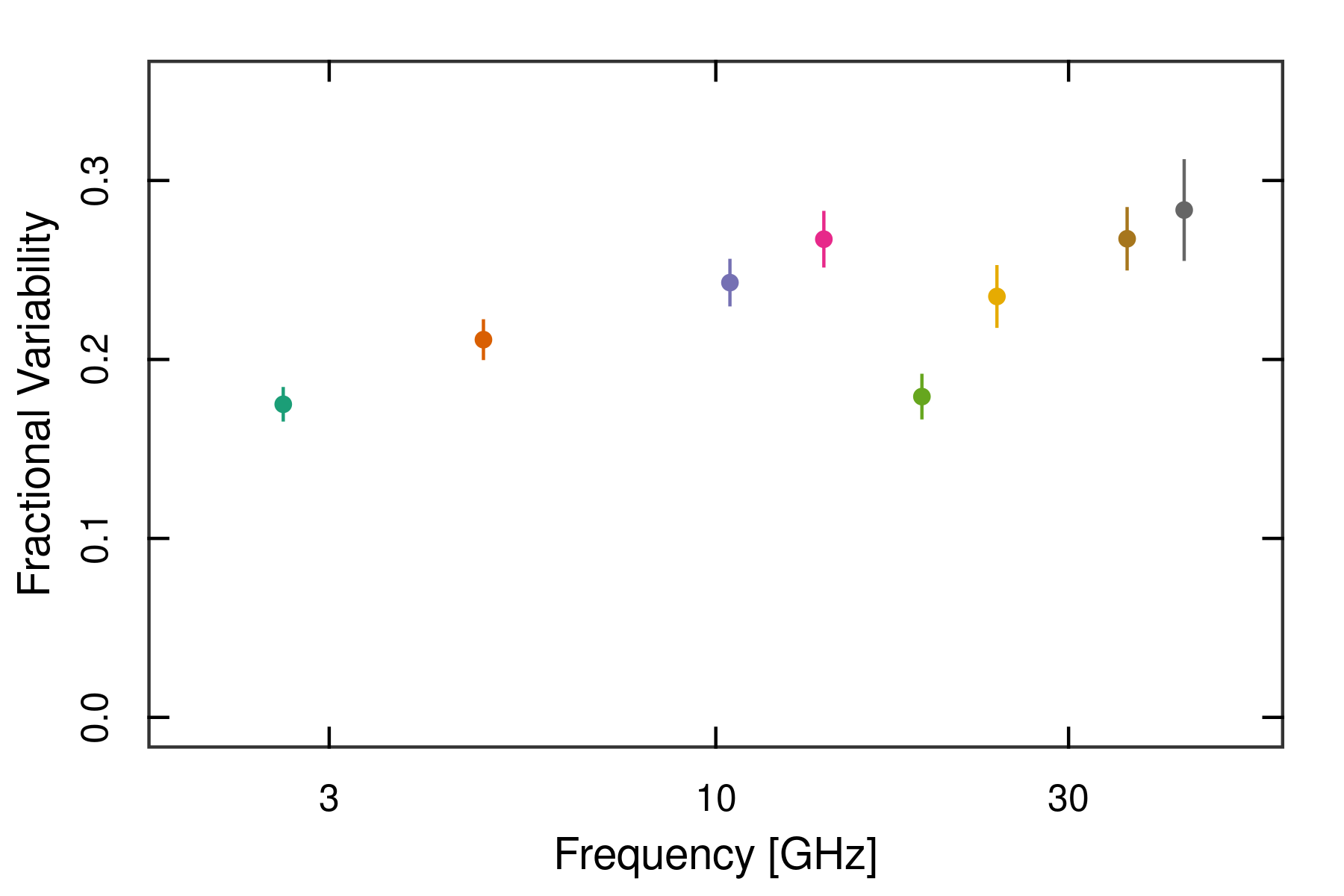}
\caption{Average fractional variability amplitude between 2.6 and 43 GHz during 2015 December to 2022 June. (Note the scarcity of data at 19 and 24 GHz during the outburst epoch 2016--2017 that partially explains the lower average value of $F_{\rm var}$ at these two frequencies).  }
    \label{fig:Fvar}
\end{figure}

Results are shown in Tab. \ref{tab:Fvar} and Fig. \ref{fig:Fvar}. In the long-term average, $F_{\rm var}$ increases from low to high frequency, between 0.17$\pm{0.01}$ (at 2.6 GHz) and 0.28$\pm{0.03}$ (at 43 GHz). Overall, $F_{\rm var}$ is highest during the large outburst in 2016--2017.  Results are further discussed below.   

\begin{table*}
%\scriptsize
	\centering
	\caption{Fractional variability amplitude $F_{\rm var}$ of the radio flux densities of OJ 287. Note that some frequencies were not or barely covered during selected years. When none or too few data points were available, no entry in the table is given; when the value is based on just few data points, the table entry is followed by `:'.  }
	\label{tab:Fvar}
	\begin{tabular}{lccccccc} 
		\hline
		 & 2016 & 2017 & 2018 & 2019 & 2020 & 2021 & 2022 \\
		\hline
2.6 GHz & $0.083\pm0.011$ & $0.059\pm0.007$ & $0.048\pm0.010$ & $0.085\pm0.013$ & $0.080\pm0.011$ & $0.105\pm0.018$ & $0.047\pm0.009$ \\
4.85 GHz & $0.148\pm0.019$ & $0.105\pm0.012$ & $0.043\pm0.008$ & $0.142\pm0.021$ & $0.094\pm0.013$ & $0.155\pm0.025$ & $0.052\pm0.010$ \\
10.45 GHz &  $0.279\pm0.035$ & $0.178\pm0.022$ & $0.053\pm0.011$ & $0.160\pm0.024$ & $0.091\pm0.014$ & $0.199\pm0.030$ & $0.064\pm0.012$ \\
14 GHz &  $0.219\pm0.041$ & $0.219\pm0.027$ & --- & $0.172\pm0.032$ & $0.068\pm0.011$ & $0.212\pm0.033$ & $0.070\pm0.013$ \\
19 GHz &  $0.079\pm0.034$: & $0.207\pm0.029$ & $0.142\pm0.029$ & $0.072\pm0.025$ & $0.070\pm0.015$ & $0.209\pm0.034$ & $0.085\pm0.017$ \\
%% 24 GHz  
32--36 GHz &  $0.253\pm0.040$ & $0.205\pm0.028$ & --- & $0.125\pm0.029$ & $0.087\pm0.020$ & $0.220\pm0.042$ & $0.096\pm0.022$ \\
43 GHz &  $0.315\pm0.062$ & $0.244\pm0.059$ & $0.211\pm0.066$ & --- & --- & $0.295\pm0.072$ & $0.139\pm0.030$ \\
		\hline
%\begin{tablenotes}
%{Notes: }
%\end{tablenotes}
	\end{tabular}
\end{table*}

\subsection{Intra-day variability (IDV)}

Two epochs were covered with daily cadence at the Effelsberg telescope: 
One in December 2015, and one in April 2017 (when for technical reasons only sources at certain sky locations could be observed, and OJ 287 was covered repeatedly as it was outbursting at that time).  
Two more epochs were observed at high cadence with the SMA in 2022 January and March. These observations allow us to search for variability within $\sim$1 d. 

Daily coverage at Effelsberg was obtained between 2017 April 6 and 16, with flux densities at 32 GHz ranging between $6.79\pm{0.04}$ Jy and $7.77\pm{0.05}$ Jy within 7 days. The most rapid change within a day was from $7.77\pm{0.05}$ Jy to $7.45\pm{0.08}$ Jy, on the order of 4$\%$ decline. 

During the near-daily cadence from 2022 January 11 to 17, SMA radio flux densities at 225.5 GHz ranged between $3.63\pm{0.18}$ Jy and $4.03\pm{0.20}$ Jy with no evidence for short-term variability within the errors. 
No short-term variability within the errors is detected with SMA between 2022 March 2 and 8, either. 

The lack of significant variability on the timescale of $\sim$ 1d beyond $\sim$4$\%$ is of interest for long-duration observations of OJ 287 on that timescale, for instance with RadioASTRON or EHT.

\subsection{Radio SEDs}

\begin{figure*}
\centering
\includegraphics[clip, trim=1.0cm 2.2cm 1.3cm 0.3cm, angle=-90, width=7.5cm, width=0.49\linewidth]{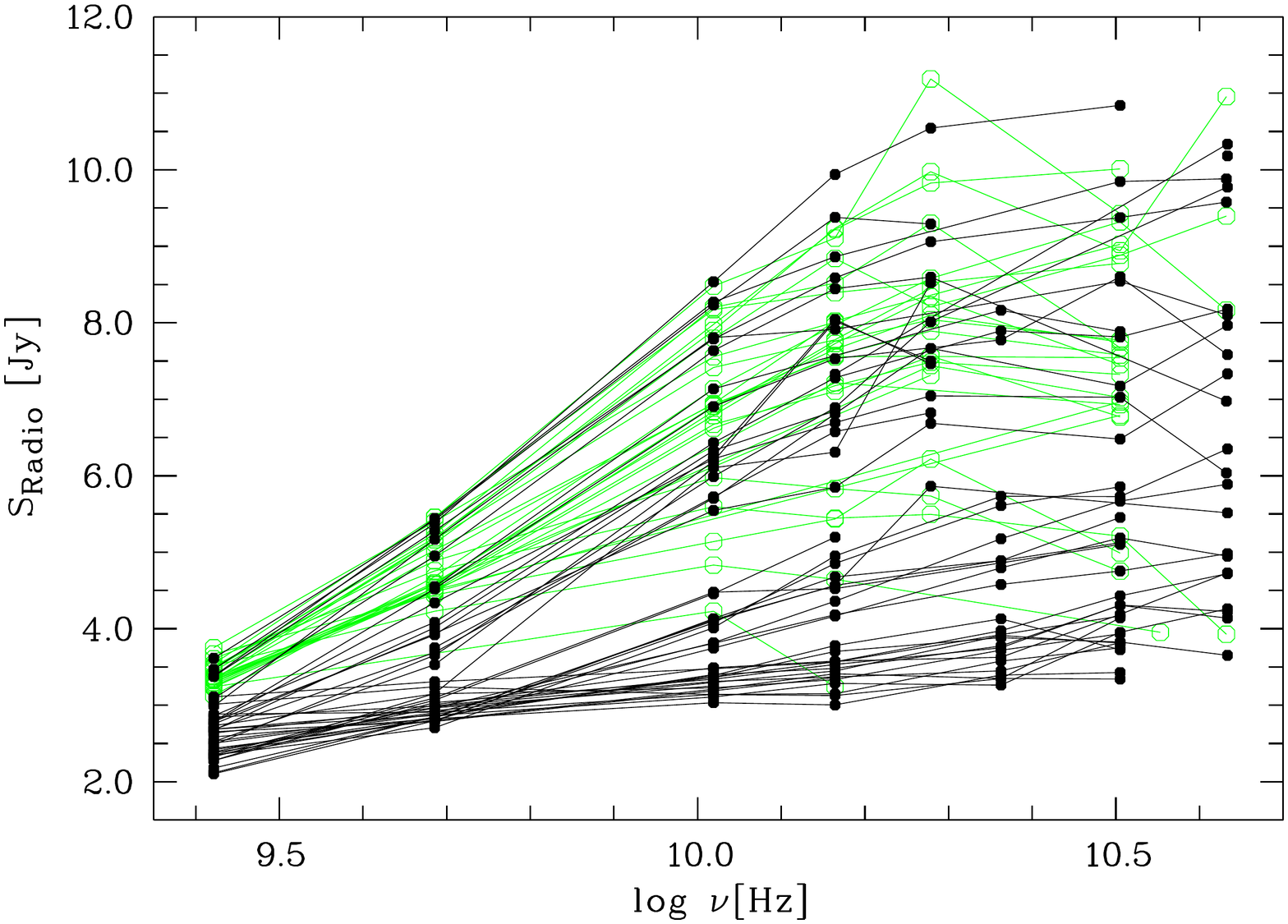}
\includegraphics[clip, trim=1.0cm 2.2cm 1.3cm 0.3cm, angle=-90, width=7.5cm, width=0.49\linewidth]{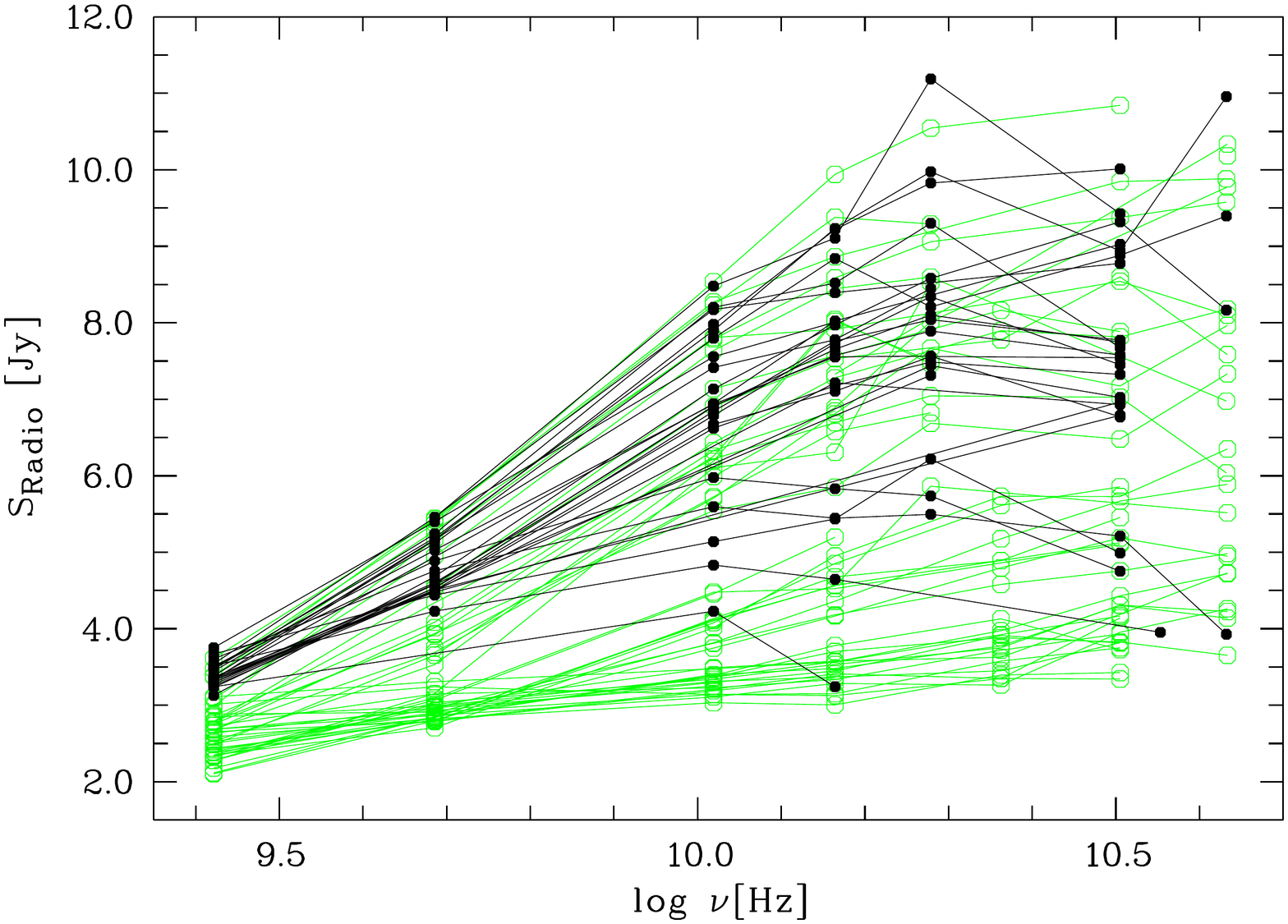}
\includegraphics[clip, trim=1.0cm 2.5cm 1.3cm 0.3cm, angle=-90, width=7.5cm, width=0.49\linewidth]{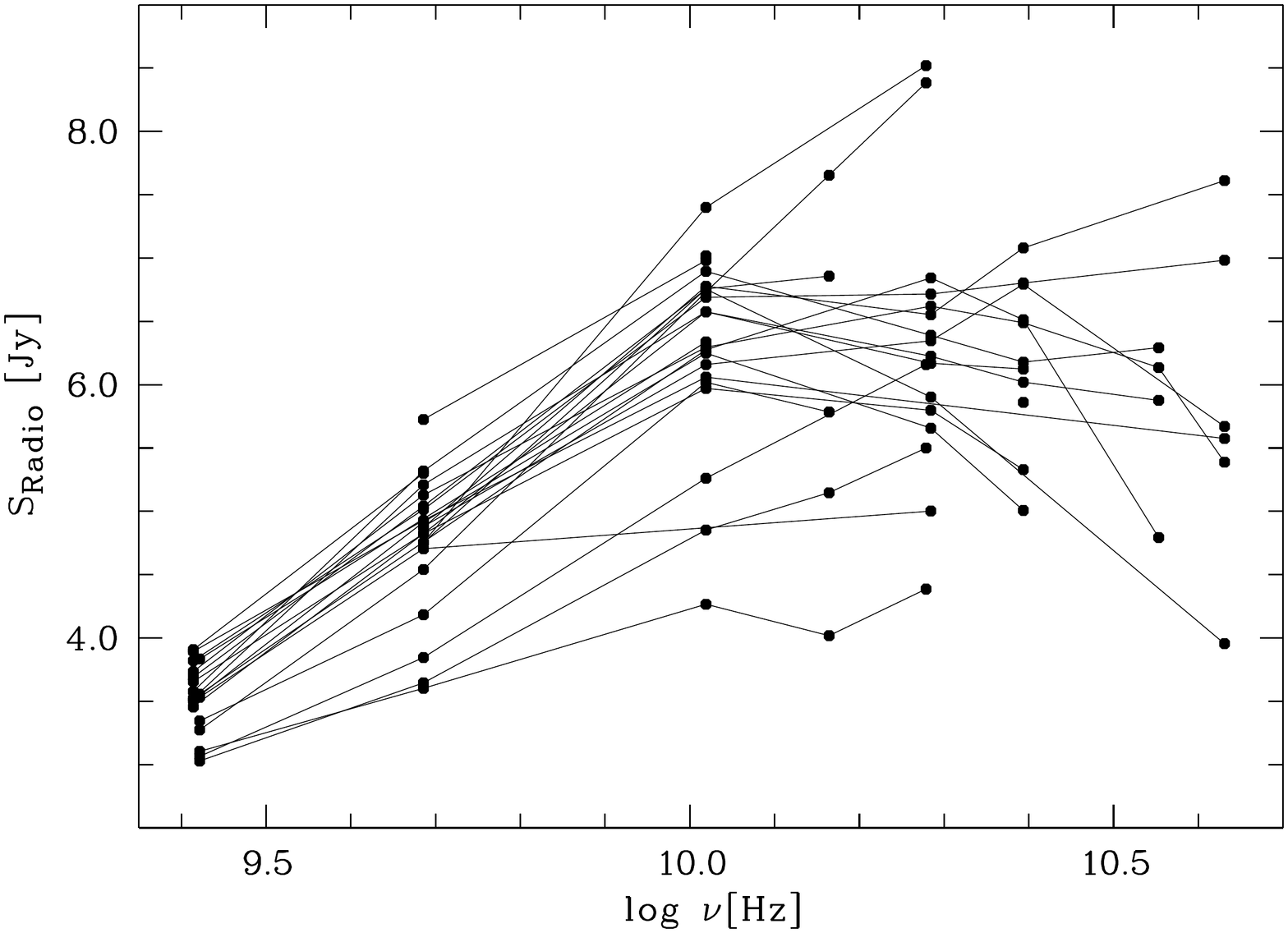}
\includegraphics[clip, trim=1.0cm 2.5cm 1.3cm 0.3cm, angle=-90, width=7.5cm, width=0.49\linewidth]{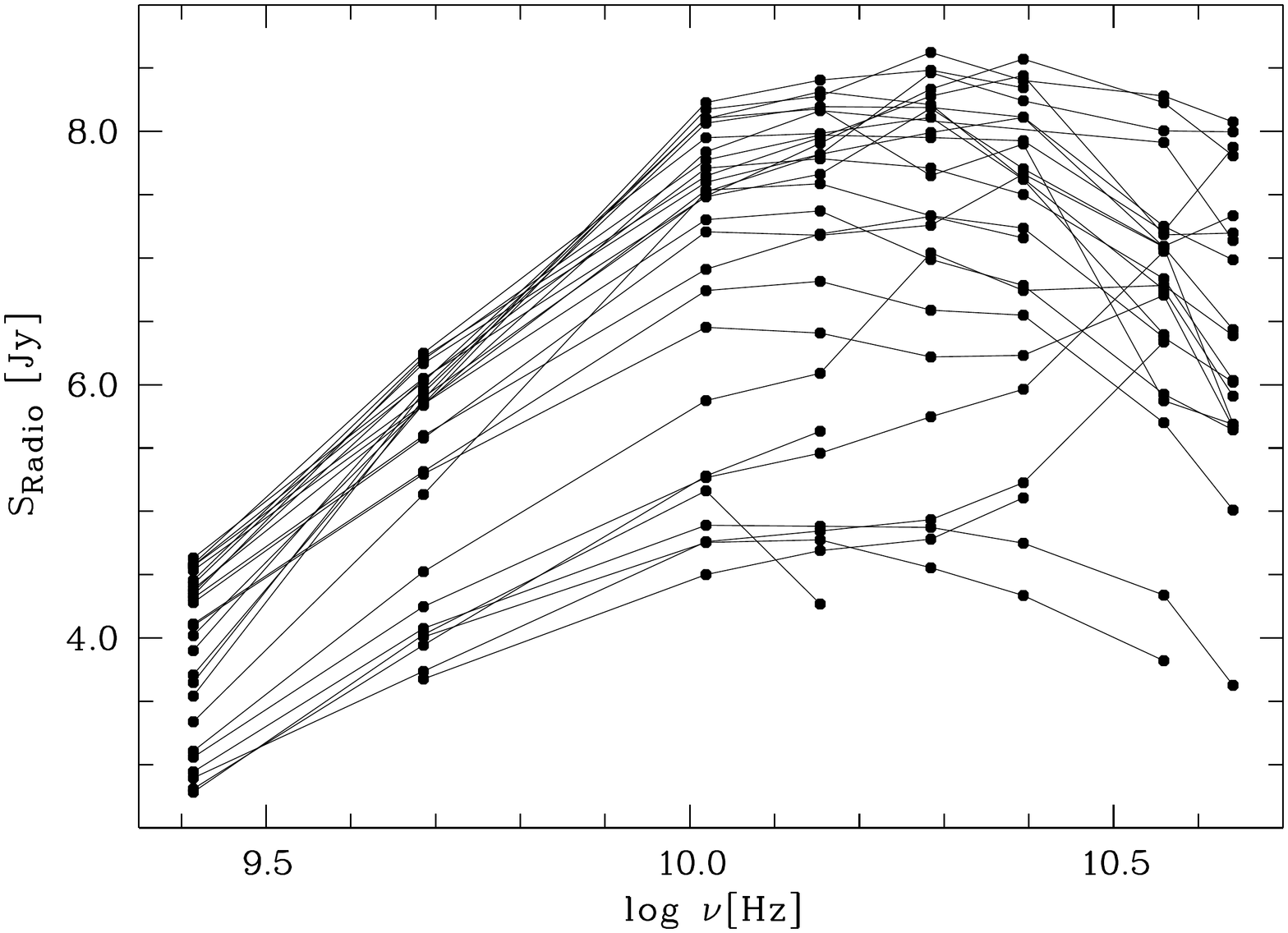}
\caption{Radio SEDs of OJ 287 between 2015 and 2022 taken at the Effelsberg telescope.
The upper two panels cover the bright 2016/2017 outburst between MJD 57816 and MJD 58062 [left panel: rise phase until the main peak in 2017 February (black; decline phase overplotted in green for comparison), right panel: decline phase until the minimum in 2017 November (black; rise phase overplotted in green)]. The bottom left panel covers the time interval 2017 November to 2019 January (exact dates: MJD 58076--58510), and the bottom right panel covers the most recent bright radio flare during the time interval 2021 May to 2022 June (exact dates: MJD 59359--59750). 
See \citet{Komossa2022a} for the time intervals MJD 58511 -- 58784, MJD 58799 -- 59028, and MJD 59034 -- 59426,  that are not repeated here.}    
     \label{fig:EB_radio_spectra}
\end{figure*}

\begin{table}
%\scriptsize
	\centering
	\caption{Observed range of spectral indices.}
	\label{tab:indices}
	\begin{tabular}{ll} 
		\hline
		Band  & $\alpha_{\nu}$ \\
		\hline
2.6--10.45 GHz & +0.03 ... +0.72 \\
10.45--36 GHz & --0.38 ... +0.35  \\
5468\AA(V)--1928\AA(W2) & --0.9 ... --1.7\\ 
		\hline
	\end{tabular}
\end{table}

The radio SEDs of OJ 287 are highly variable. Turn-over frequencies
range between 10 and 36 GHz, or are outside the observed frequency band at $>$43 GHz at other epochs.  The bright 2016/2017 outburst shows a strong spectral evolution. 

The detailed shapes of all SEDs are displayed in Fig. \ref{fig:EB_radio_spectra} (with the exception of SED sequences between 2019 and 2022 that were already presented in \citet{Komossa2022a} and are not repeated here. Those cover the epochs MJD 58511 – 58784, MJD 58799 – 59028, and
MJD 59034 – 59426). The time period 2016 was already in our first paper on radio results obtained within the MOMO project   \citep{Myserlis2018} and is shown here again to  visualize the spectral evolution of the outburst that continued into 2017.   

In addition, power-law spectral indices were calculated at selected frequencies between 2.6 and 36 GHz (Tab. \ref{tab:indices}).  
The observed spectral index $\alpha_{\nu}$ is defined as 
$S_\nu \propto \nu^{+\alpha_\nu}$. 
Even though highly variable, spectra remain in the flat-spectrum regime ($\alpha_{\nu} > -0.5$) or are inverted.   
Spectral indices
range between $\alpha_{\nu,2.6-10.45}$ = 0.03 and 0.72, and $\alpha_{\nu,10.45-36}$ = --0.38 and 0.35.
SMA observations are generally not simultaneous within a day with the Effelsberg observations. 
We have computed some representative spectral indices at high- or low-states, and when observations were carried out within 0--7 days of each other, and find 
$\alpha_{\nu,32-226.9}$ = --0.18 (SMA low-state, 2016 Jan. 30), $\alpha_{\nu,32-225.5}$ = --0.21 (SMA high-state, 2017 Feb. 16), and $\alpha_{\nu,36.25-225.5}$ = --0.31 (SMA high-state, 2021 Nov. 11).

\begin{table}
%\scriptsize
	\centering
	\caption{Time lags with respect to the flux density at 10.45 GHz. Negative values mean $S_{\rm 10.45}$ is leading. }
 %% Negative values mean $S_{\rm{10.45}}$ is  leading, positive values mean $S{_\rm{10.45}}$ is lagging. }
	\label{tab:DCF}
	\begin{tabular}{lr} 
		\hline
		$\nu$  & Lag~~~~~~~ \\
		\hline
2.6 GHz & $-51^{+17}_{-18}$ d \\
4.85 GHz & $-18^{+10}_{-10}$ d\\
10.45 GHz &  --- \\
14 GHz &  $2^{+6}_{-7}$ d \\
19 GHz &  $-1^{+7}_{-8}$ d \\ 
36 GHz &  $16^{+7}_{-7}$ d \\
		\hline
%\begin{tablenotes}
%{Notes: }
%\end{tablenotes}
	\end{tabular}
\end{table}

\begin{figure*}
\includegraphics[width=0.49\linewidth]{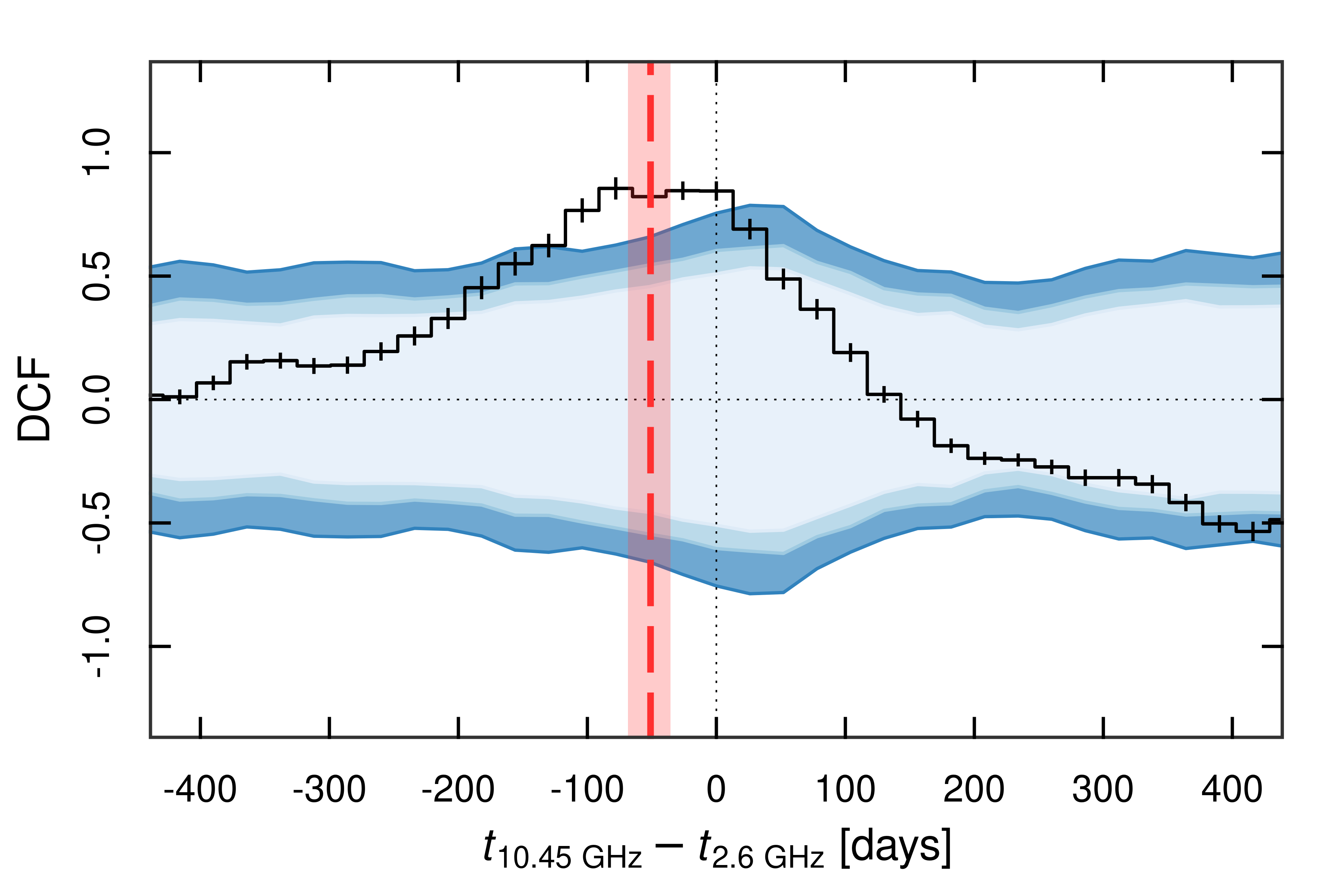}
\includegraphics[width=0.49\linewidth]{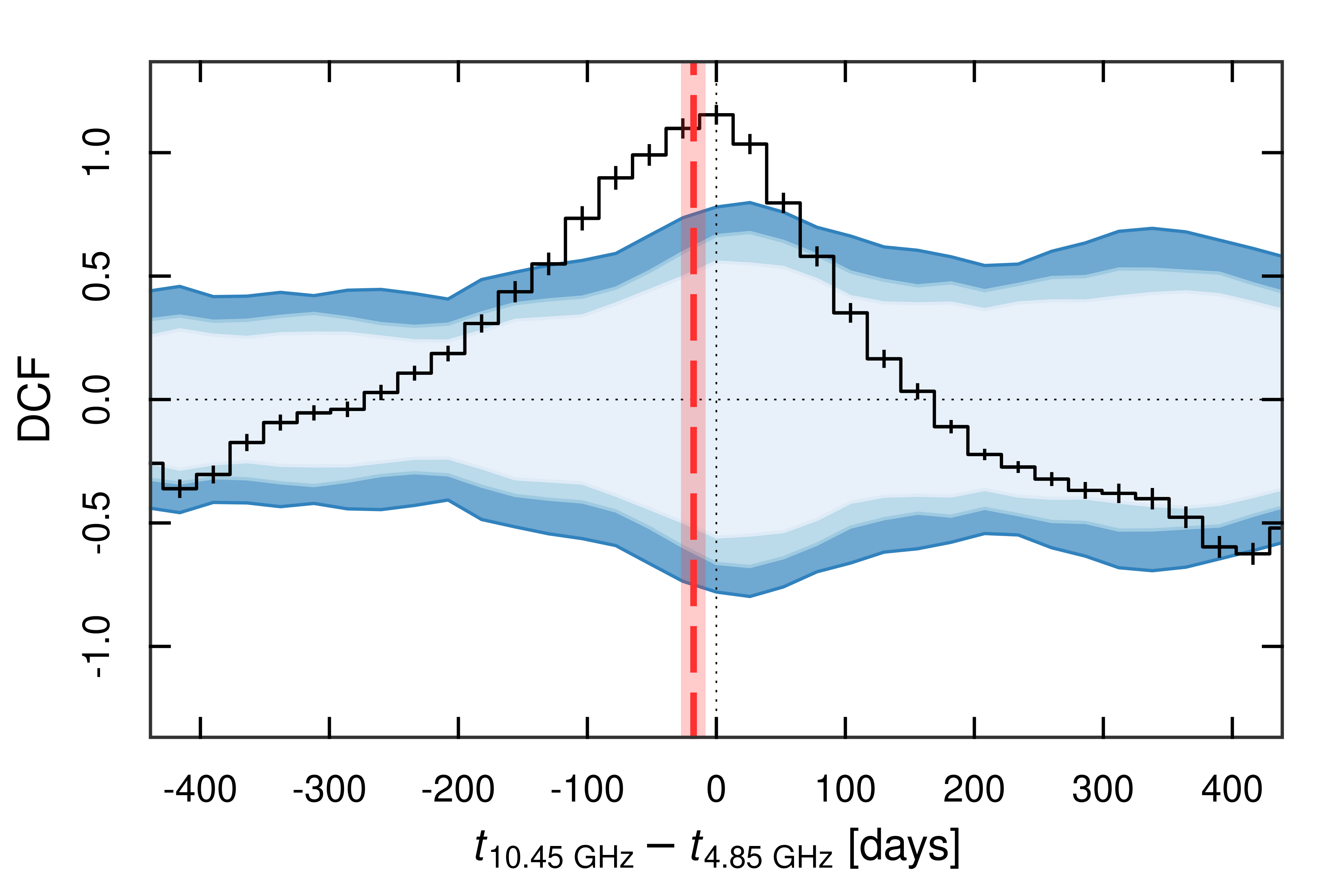}
\includegraphics[width=0.49\linewidth]{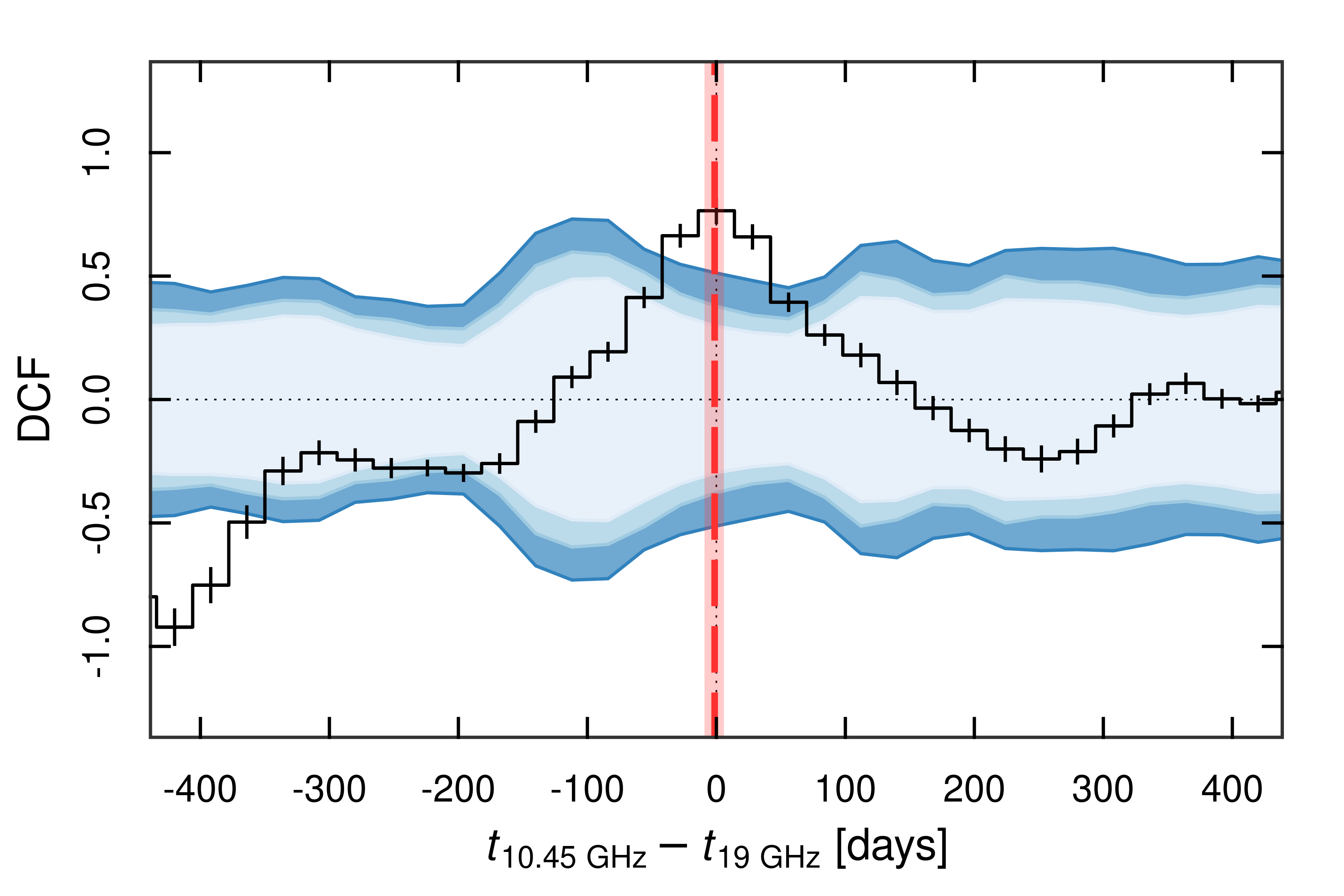}
\includegraphics[width=0.49\linewidth]{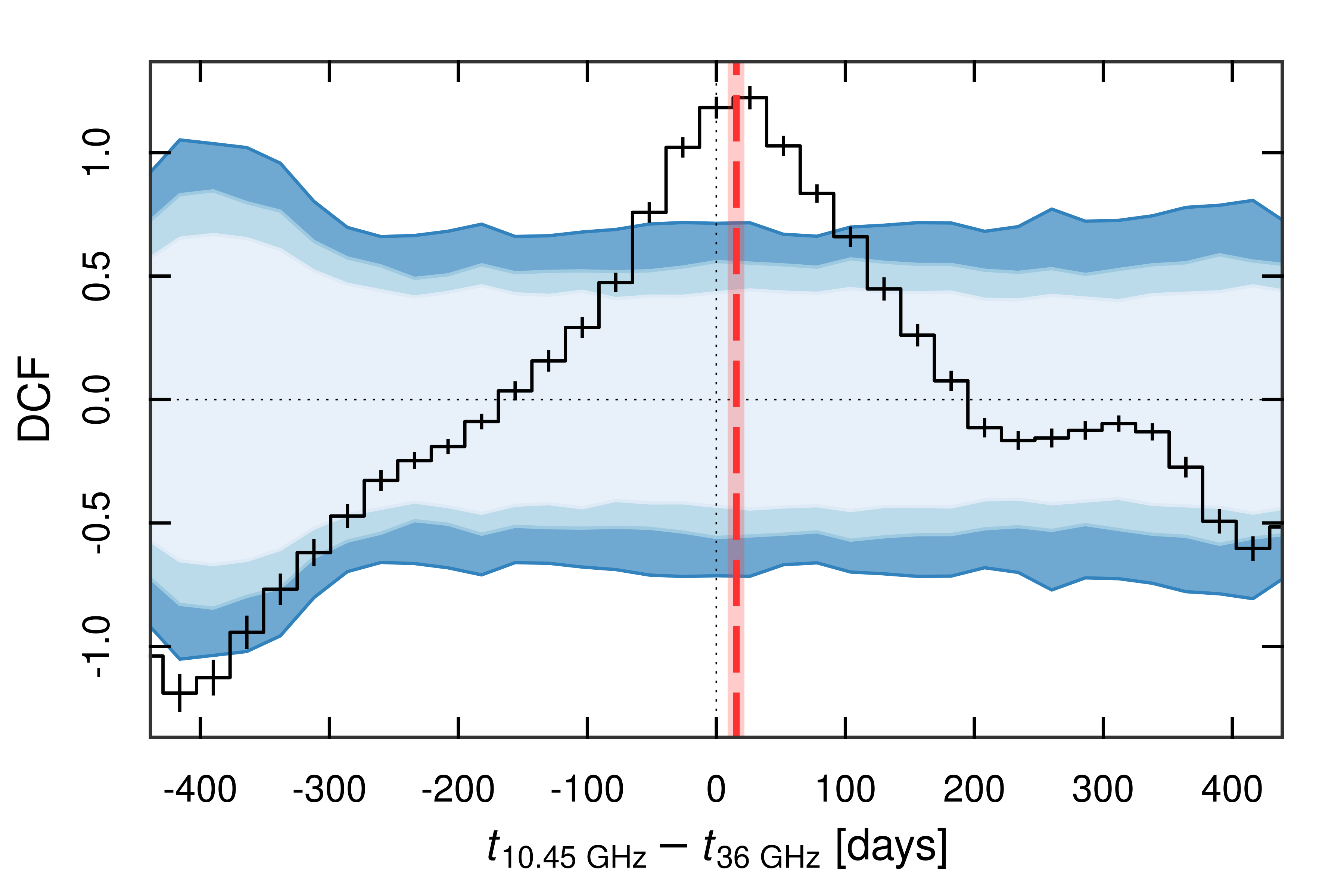}
\caption{ 
The $S_{\rm 10.45 GHz} - S_{\nu}$ DCF (black solid line) was computed for each flux density between $\nu$=2.6 and 36 GHz
as labelled in each panel. Filled regions indicate the $\pm90^{\mathrm{th}}$ (light blue), $\pm95^{\mathrm{th}}$ (medium blue), and $\pm99^{\mathrm{th}}$ (dark blue) percentiles from the $N=10^3$ light-curve simulations. Horizontal dotted lines indicate $\mathrm{DCF}=0$, 
% mark zero correlation
and vertical dotted lines mark $\tau = 0~\mathrm{days}$. 
Negative time-axis values indicate $S_{\rm 10.45 GHz}$ leading the other band, positive values indicate lagging. The vertical red line indicates the measured time lag, with filled regions corresponding to the $68~\mathrm{per~cent}$ confidence interval of the lag.
} 
\label{fig:DCF-opt-UV}
\end{figure*}

\subsection{Time delays and DCF}

Opacity effects due to synchrotron self-absorption 
lead to delays in the rise times 
of flares \citep[e.g.,][]{Lee2020}.
In order to quantify this effect  
across the long-term light curve at all frequencies between 2015 December and 2022 June 
we have computed the discrete correlation function (DCF).  
The DCF technique was designed to analyze unevenly sampled data sets 
\citep{Edelson1988}. 
We employ the DCF to search for time lags between the radio bands, measured against the 10.45 GHz data set.

We computed the DCFs as prescribed in \citet{Edelson1988}, selecting the time step, $\tau$, over which the DCFs were computed, to be twice the median time step across the entire light curve at each radio frequency. 

To evaluate the significance level of measured lags we produced confidence contours for each DCF by simulating $N=10^3$ artificial light curves in each band, following the prescription of \citet{Timmer1995}, assuming a power spectral density (PSD) of $P\left(f\right) \propto f^{-\alpha} = f^{-1.97}$. This value of $\alpha$ is based on the results of a structure function (SF) analysis \citep[following][]{Gallo2018} of the 10.45 GHz light curve,
that yielded an index of $\beta=0.97$ for timescales of $\lesssim240~\mathrm{days}$, and subsequently taking $\alpha = \beta + 1$.

The artificial light curves were then used to compute artificial DCFs with the band-of-interest light curve, allowing for the computation of $90^{\mathrm{th}}$, $95^{\mathrm{th}}$, and $99^{\mathrm{th}}$ percentiles based on the distribution of 
artificial DCFs at each time step.  

\begin{figure}
\centering
\includegraphics[clip, width=\columnwidth]{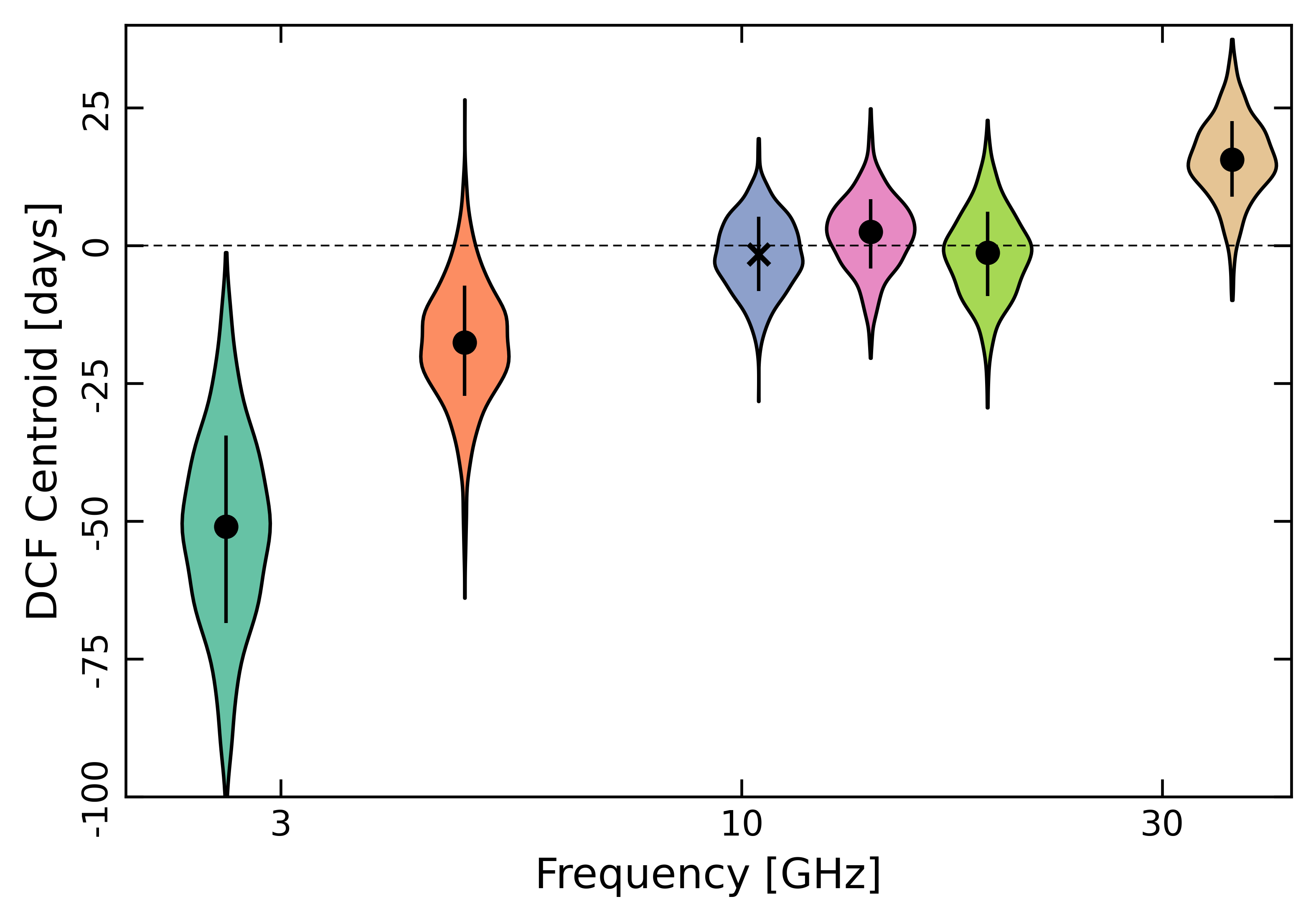}
\caption{DCF lag versus radio frequency. The violin-shaped regions represent the distribution of the centroid measurements at each frequency. The reference band is 10.45 GHz.}
    \label{fig:DCF_interpol}
\end{figure}

To evaluate the error on the measured lags, we applied a least-squares minimization of a Gaussian function to the central peak, defined as the points with correlation strength corresponding to $\geq 50\%$ of the peak value,  within $\tau=\pm400~\mathrm{days}$, which enabled measurement of the DCF centroid. To evaluate the $68~\mathrm{per~cent}$ confidence region of the lag (centroid) measurement, we employed the resampling technique outlined by \citet{Peterson1998}.
Briefly, for each point in the two light curves used in a given DCF the flux density is resampled according to a Gaussian distribution with mean and standard deviation equal to the flux density and its error for that point. Then, a random subset of points from each light curve is selected and used to compute a new `resampled' DCF, from which the centroid is measured in the aforementioned way. This procedure is conducted $N=10^3$ times, allowing the $68~\mathrm{per~cent}$ confidence interval on the `true' DCF centroid to be evaluated using the $16^{\mathrm{th}}$ and $84^{\mathrm{th}}$ percentiles of the `resampled' DCF centroid distribution.
Results are shown in Fig. \ref{fig:DCF-opt-UV} and \ref{fig:DCF_interpol}, and Tab. \ref{tab:DCF}.

In a next step, we have re-run all analyses on the two brightest outbursts of 2016/2017 and 2021/2022 only, in the time interval MJD 57370--58077 and MJD 59426--59751, respectively. Overall, the two events show similar time lags. Longer opacity delays at low frequencies are found for the 2016/2017 outburst with $\delta t_{\rm}$=--83$^{+11}_{-12}$ days (2.6 GHz) and -38$\pm{9}$ days (4.85 GHz). 

\begin{figure*}
\centering
\includegraphics[clip, trim=0.9cm 5.3cm 1.0cm 2.3cm, width=16cm]{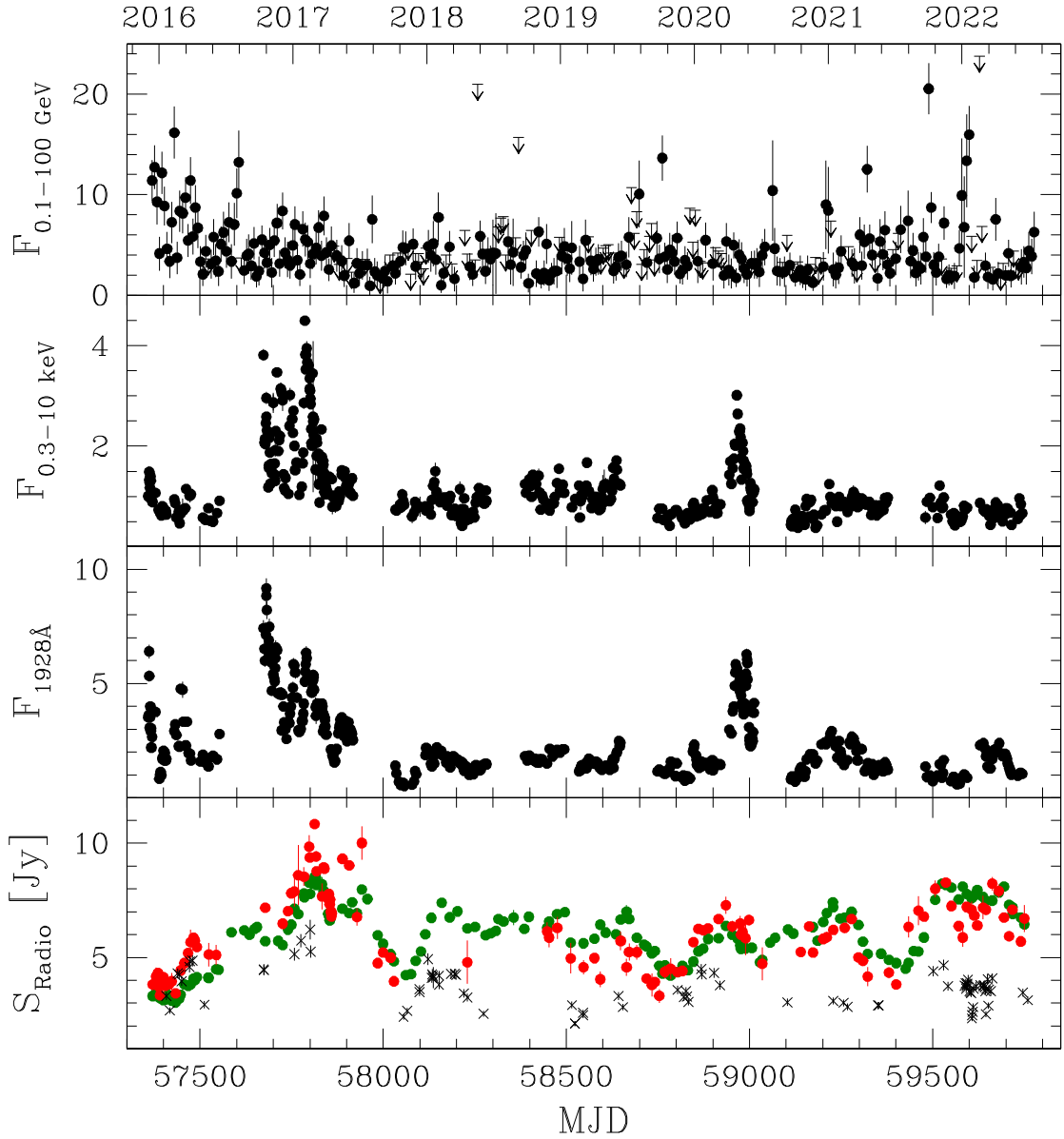}
    \caption{MWL flux light curve of OJ 287 between 2015 December and 2022 June. From top to bottom: Fermi $\gamma$-rays, Swift X-rays, Swift UV (at 1928\AA), and Effelsberg and SMA radio observations at selected frequencies (green circles: 10.45 GHz, red circles: 32-36.25 GHz, black crosses: SMA data between 213 and 351 GHz).
    The $\gamma$-ray flux (observed 0.1--100 GeV band; one-week averages), the absorption-corrected X-ray flux (observed 0.3--10 keV band), and the extinction-corrected UV flux at $\lambda_{\rm obs}$=1928\AA~are given in units of 10$^{-11}$ erg s$^{-1}$ cm$^{-2}$. The radio data are reported as flux densities in Jy. 
    The gap in Swift  observations between June to September each year is due to the close proximity of OJ 287 to the Sun, such that it becomes unobservable with Swift. Fermi and Effelsberg observations are not affected by this Sun constraint. 
   Note that the single brightest $\gamma$-ray data point from
   2015 December is off the scale (see Fig. \ref{fig:MWL-2017} instead). 
   }
    \label{fig:MWL-all}
\end{figure*}

\section{MWL light curves}

MWL light curves [Fermi $\gamma$-rays, Swift X-rays, Swift UV-W2 (representative for the 3 optical and 3 UV bands that are always strongly correlated, even though the spectral index does change)] are shown in Figs. \ref{fig:MWL-all}, \ref{fig:MWL-2017}, \ref{fig:MWL-2022} (see \citet{Komossa2021c} for the full Swift long-term light curve including all optical and UV bands between 2005 and 2021 March). 

The large 2016/2017 X-ray--UV--optical outburst \citep{Komossa2017, Komossa2020} is well correlated with the radio emission that reaches its highest flux density at high frequencies during the whole MOMO monitoring period (Fig. \ref{fig:MWL-2017}). The X-ray--optical outburst shows some pronounced substructure with several peaks seen at all wavebands, and the optical--UV bands reached their highest peak flux earlier than the X-ray band. The brightest optical--UV peak was reached on 2016 October 20, whereas 
the brightest X-ray peak was only reached on 2017 February 2. 
Our monitoring cadence at that time was 1 day. 
The brightest radio peak was observed close in time to the brightest X-ray peak, on 2017 February 28 (MJD  57812$^{+4}_{-12}$ days, where the uncertainty reflects the cadence of the radio monitoring at that epoch). 
A second, earlier radio peak detected at all radio frequencies occurred on 2016 July 31 (MJD 57600$^{+34}_{-7}$ days), at an epoch when OJ 287 was in Swift Sun constraint, and no data in the optical--UV and X-rays were taken. 

Fig. \ref{fig:spectral-indices} compares the spectral indices $\alpha_{\nu}$ measured in selected radio bands with the optical--UV spectral index.
The optical--UV spectral index, measured between V (5468\AA) and UVW2 (1928\AA), shows a `bluer-when-brightest' behavior, and is steepest during several low-states. Optical and radio spectral indices are correlated. 

\section{Discussion} 

\subsection{Radio variability}

The radio light curves of OJ 287 show high-amplitude variability with multiple flares and several deep low-states.  
On average, the data show an increase in $F_{\rm var}$ from low- to high frequencies (Fig. \ref{fig:Fvar}), as expected in a plasma that becomes increasingly optically thin towards higher radio frequencies.  An exception is seen around 19 GHz, where $F_{\rm var}$ shows a dip. However, the bulk of the effect is caused by
the lower number of 19 GHz observations at the beginning of the monitoring interval in 2016, when a strong rise into high-state occurred, and $F_{\rm var}$ was highest.

Fractional variability amplitudes in the radio band are found to be lower than values obtained for the UV--optical and X-ray light curves ($F_{\rm var,V}=0.3$ , $F_{\rm var,X}=0.4$ respectively) at similar time intervals \citep{Komossa2021c}, reflecting decreasing opacity effects in the optical regime.  The trend across frequency and the values of $F_{\rm var}$ of OJ 287 are similar to those observed in other blazars \citep[e.g.,][]{Schleicher2019}. 
Values of $F_{\rm var}$  are highest during the 2016/2017 outburst.
This outburst will be further discussed below.   

The majority of the radio and optical flares are closely correlated, with the optical leading the radio \citep[see also][]{Komossa2022a}. This hints at a co-spatial origin of optical and radio emission.
An exception is the latest 2021--2022 radio flare.  Its bright state in late 2021  
is not accompanied by major optical flaring. 
However, it is  
associated with two sharp $\gamma$-ray flares;
the two brightest in recent years (Fig. \ref{fig:MWL-2022}). One at the end of 2021 October and a second, broader one that starts in late December.
This type of behaviour indicates a possible causal connection.   
It is conceivable that the first sharp $\gamma$-ray flare occurs 
when the jet is hitting the first recollimation shock. The second $\gamma$-ray flare is broader/slower, indicating a larger size of the emission region. It is possibly produced further downstream where the jet also becomes broader.  
We also note that the 2016/2017 outburst is accompanied in its rise phase by a $\gamma$-ray flare in 2016 August. 

\subsection{Deep fades every 1--2 years}

A remarkable sharp, symmetric, UV--optical deep  fade was detected with Swift during 2017 October--December \citep{Komossa2021c} that we now also cover in the radio band where it is seen as well at all frequencies. Similar, but less pronounced, UV--optical low-states are also detected in our light curves in late 2019, late 2020, and late 2021.  
Several of these have counterparts in the radio band in the form of deep dips of the radio emission at all bands, seen in late 2015, late 2017, and late 2019, whereas the radio deep fade in 2021 comes a few months earlier, in mid 2021.
Another such radio dip is seen in the radio light curve of \citet{Lee2020} in late 2013.

With only a few well-documented occurrences in our data, a rigorous period search is not yet warranted. However, we note that the 
autocorrelation function (ACF) at 10.45 GHz (Fig. \ref{fig:ACF}) does show evidence for periodicity at 410 days and at 650 days. It is also interesting to note that several previous studies of OJ 287 reported evidence for periodicities in the range of 1--1.7 yrs in the radio
\citep{Hughes1998, Hovatta2008, Britzen2018}, and $\sim$400 days in the optical \citep{Bhatta2016,Sandrinelli2016}.
Independent of the possibility of semi-periodicity, the question is raised what causes these deep fades.  They may just represent the quiescent states inbetween epochs of flaring. However, their symmetric nature of fading and rising without remaining in quiescent states for significant amounts of time suggests otherwise. Deep fades could be caused by extreme absorption events, de-beaming due to de-collimation or wobbling of the jet, or they could be
directly related to the electron launching process at accretion-disk scales in the first place. 

The deep fades do not represent SAV \citep[Symmetric Achromatic Variability;][]{Vedantham2017} events that were speculated to be caused by milli-lensing from intermediate-mass black holes, since they are not achromatic but show a systematic change in radio spectral indices (Fig. \ref{fig:spectral-indices}). Some deep fades show a U-shaped structure that is particularly similar to the SAVs (e.g., the 2017 UV--optical deep fade or the 2021 radio deep fade), but since they go deeper than the surrounding quiescent level and are not achromatic, the observations of OJ 287 also show, that there are mechanisms other than possible milli-lensing to produce these characteristic structures in blazar lightcurves.    

Independent of the interpretation of these events, the prediction of their future occurrences is of interest for deep imaging studies of the host galaxy of OJ 287 \citep[as done shortly after the 2017 optical deep fade by][]{Nilsson2020} 
with JWST, the Hubble Space Telescope, or ESO's Enhanced Resolution Imaging Spectrograph (ERIS), and for high-resolution optical spectroscopy in search of host absorption features for direct SMBH mass measurements from stellar velocity dispersion $\sigma_*$. At epochs of deep fades the blazar glare itself is least interfering with the measurements.   

\subsection{Absence of predicted 2021 December precursor flare activity}

\begin{figure*}
\centering
\includegraphics[clip, trim=0.9cm 5.3cm 1.0cm 2.3cm, width=16cm]{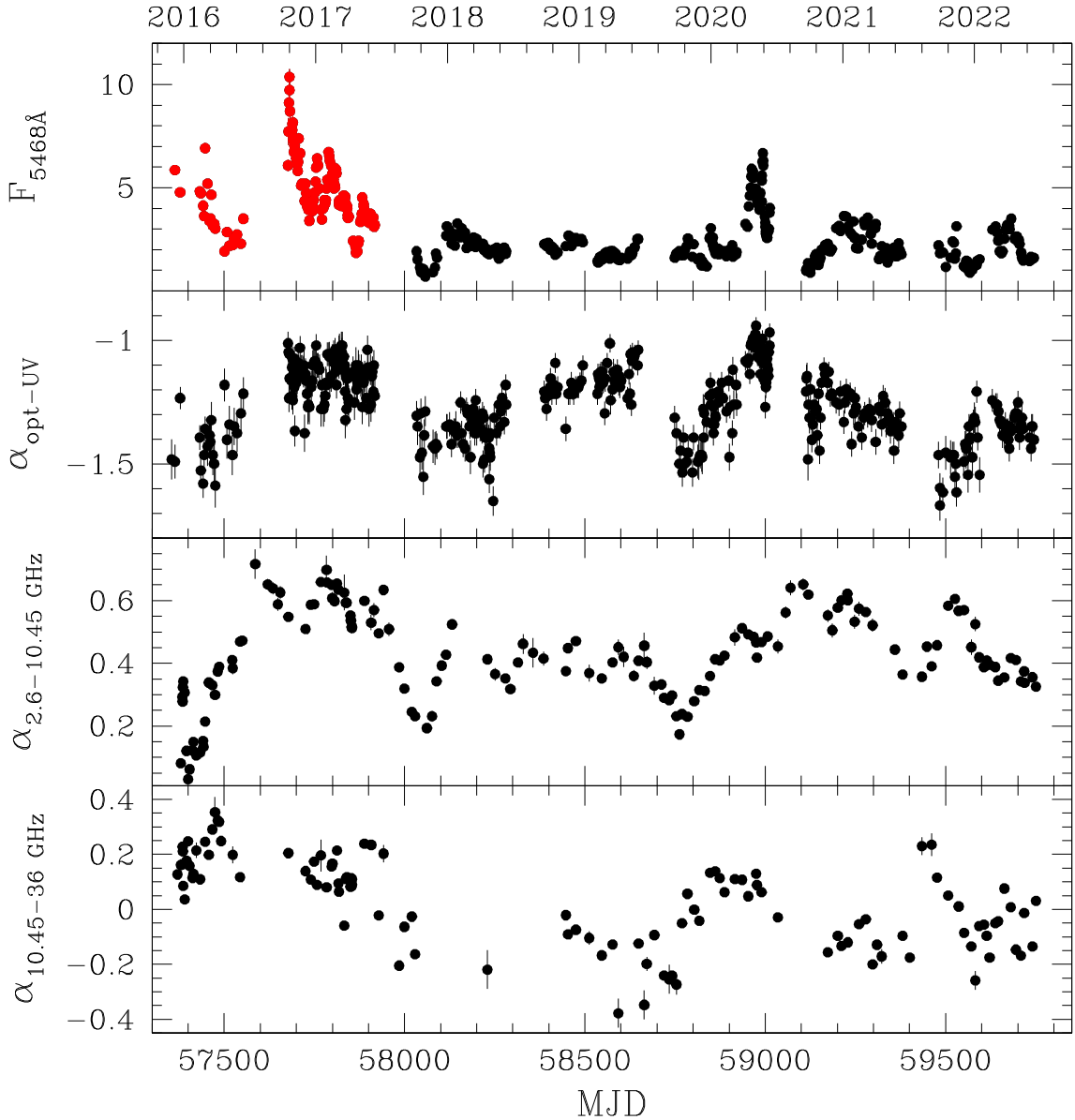}
% {fig_oj287_a_uv_a_radio_lc}
    \caption{Spectral indices $\alpha_{\nu}$ in selected optical and radio bands (lower panels; $S_\nu \propto \nu^{\alpha_\nu}$), and the Swift V flux at 5468\AA ~in units of 10$^{-11}$ erg s$^{-1}$ cm$^{-2}$ for reference (upper panel). 
    The optical--UV spectral index is measured between V (5468\AA) and UVW2 (1928\AA). The double-peaked 2016/2017 outburst is marked in red. The first maximum is reached  in 2016 February, the second and brighter one in 2016 October.  
    }
    \label{fig:spectral-indices}
\end{figure*}

It was speculated that one of the optical flares in the 2005 light curve of OJ 287 was driven by binary SMBH activity in the form of a `precursor flare' preceding the main outburst.  Based on the PB model it was predicted that the precursor flare would repeat in 2021, on December 23, with an optical thermal bremsstrahlung spectrum of index  $\alpha_{\nu}=-0.2$ (and without any X-ray or radio counterpart; \citet{Valtonen2021}, their Section 3). However, OJ 287 was found in a deep optical--UV low-state throughout December 2021, 
and any meaningful optical--UV flaring activity at that time period could be excluded
\citep{Komossa2022b}. 
Further, neither the rise in emission after the deep low-state, nor any other UV--optical activity until June 2022, showed the predicted thermal bremsstrahlung spectrum. 
Observed spectral indices $\alpha_{\nu, \rm{opt-UV}}$ 
range between --1.2 and --1.54 (Fig. \ref{fig:spectral-indices}) whereas the predicted thermal bremsstrahlung spectrum would have had $\alpha_\nu \sim -0.2$.  
We therefore conclude that the bremsstrahlung flare predicted by the PB model did not happen, neither in 2021 December nor at any other epoch. 

We now turn our attention to the remarkable 2016/2017 outburst and its  interpretation in the context of previous variants of binary models that do not involve a strong orbital precession, in contrast to the PB model. 

\subsection{The 2016/2017 outburst, its interpretation as the latest double-peaked outburst of OJ 287, and implications for binary SMBH models}

\subsubsection{MWL properties of the 2016/2017 outburst}

A high-amplitude outburst occurred in 2016/2017,  detected in the course of our Swift monitoring in all bands from the X-rays to the optical with a systematic X-ray `softer-when-brighter' variability pattern with steep X-ray spectral indices $\Gamma_{\rm x} \simeq$ 3 at peak brightness
\citep{Komossa2017, Komossa2020}. 
While \citet{Komossa2017} first speculated about a thermal nature of this outburst, detailed follow-ups then established the nonthermal, Synchrotron, nature of this event, based on the following arguments: 
First, with Swift, we detected X-ray
flux doubling timescales as short as 4 days, shorter than the light-crossing time at the last stable orbit of the accretion disk around a  primary SMBH if its mass was $M_{\rm BH} = 1.8 \times 10^{10}$ M$_{\odot}$ as required in the PB model, ruling out a primary's disk origin \citep{Komossa2021a}{\footnote{The argument only holds within that specific binary model. If the SMBH mass of OJ 287 is only around $10^{8}$ M$_{\odot}$, then the most rapid variability timescales would still be consistent with a disk origin, and this particular argument breaks down. }}. 
Second, optical--UV DCF results \citep{Komossa2021c} are consistent with synchrotron theory, but the lags are
too small for accretion–disk reverberation \citep{Kammoun2021} of a SMBH with a mass as low as $\sim10^8$ $M_{\odot}$. 
Third, the Swift X-ray spectra are well explained by a soft synchrotron emission component and show the same softer-when-brighter variability pattern also seen during other outbursts \citep{Komossa2020, Komossa2022a, Idesawa1997}.
Fourth, the early phase of the outburst was already associated with an increase in radio emission \citep{Myserlis2018, Kapanadze2018, Lee2020}, 
the levels of radio polarization were high \citep{Myserlis2018, Goddi2021}, and VHE emission near the X-ray peak was detected by VERITAS 
\citep{OBrien2017} [only the Fermi $\gamma$-ray regime lacked significant flaring \citep{Kapanadze2018, Komossa2022a}].   

Here, we provide the coverage of the whole outburst at multiple radio frequencies between 2.6 and 225 GHz,
strongly re-confirming the nonthermal nature of the outburst. The radio emission is closely correlated with the MWL emission, and the brightest radio and X-ray peak coincide closely
and are both reached in 2017 February (Fig. \ref{fig:MWL-2017}). 
The radio peak flux in 2017 is among the highest so far recorded from OJ 287. It is comparable to the brightest state in the 1972--1996 light curve reported by \citet{Valtaoja2000}. 

In the next Section, we will argue that the extraordinary 2016/2017 outburst matches the properties of the previous famous semi-periodic double-peaked outbursts of OJ 287 of the 1970s -- 2000s and that it is the latest such event in the sequence. 
First assuming strict periodicity of the previous double peaks with $\Delta t = 11.86$ yr and using the timing of the well-observed 1994 burst
\citep{Valtaoja2000}, the latest one should have occurred in 2018. However, no burst was observed at that epoch, but in 2016/2017 instead.  
Therefore, next, we drop the requirement of strict periodicity as also done in most previous works and for the additional reasons given below, and we then discuss implications for binary SMBH models of OJ 287. 

\subsubsection{Re-interpretation of the 2016/2017 outburst}

The fact that the precursor flare predicted by the PB model was not observed, and the fact that the most pronounced outburst in recent years was the one in 2016/2017 (not expected in the PB model), leads us to re-consider alternative binary SMBH models of OJ 287. We ask if the timing of the 2016/2017 outburst fits any of the previous models, that involve less extreme orbital precession and that lead to the expectation of more regularly-spaced outbursts than the PB model, with a known variance on the order of 1--2 years as directly observed previously during the major double-peaked outbursts.
Most authors reported time intervals of $\Delta t$=11--12 yrs between the outbursts \citep[e.g.,][]{Sillanpaa1988, Valtaoja2000} directly
from observations,  with additional uncertainties in $\Delta t$ of at least several months based on
(1) gaps in the cadence of observations during the 
epochs of dedicated monitoring since 1970, (2) additional gaps of $\sim$3 months each year when OJ 287 was
unobservable due to its solar proximity,  
(3) the increasing availability of photographic plate archives that cover selected time
intervals since the 1880s (so that the timing of single peaks could shift back or forth with the availability of new data sets), 
(4) each authors' preference for the reliability of one or another time period of photographic plate data,
and/or (5) different methods of single peak determinations based on the same data sets over time.

If we take the timing of the first peak in the 1970s when the dedicated monitoring started, $t_{\rm peak}$=1971.13 \citep{Valtonen2021}, and the peak of the first 2016 V-band flare at 2016 February 29 (Fig. \ref{fig:spectral-indices}; upper panel), then this results in an average $\Delta t$ of 11.3 yr for 4 cycles. 

How do the properties of the 2016/2017 
outburst compare with those of previous double-peaked outbursts of OJ 287 ? 
First, the 2016/2017 outburst shows a second, bright optical peak on 2016 October 20 (Fig. \ref{fig:MWL-2017}), 
8 months after the first peak, matching the well-established characteristic historical double-peak structure of past flares, with the second peak appearing $\approx$1 yr after the first peak 
\citep[e.g.,][]{Valtaoja2000}. 
Further, the 2016/2017 outburst agrees with the  observation of \citet{Valtaoja2000} in that the second peak is associated with  radio flaring. However, we also find evidence that the 
first peak has a  radio counterpart at high frequencies. It is possible that it was missed in earlier observations because it was self-absorbed 
or due to gaps in the coverage.  
The whole 2016/2017 outburst is associated with radio emission. 
Another similarity of the 2016/2017 outburst with previous double-peaked outbursts is the fact that no major X-ray flare is associated with the first peak \citep{Idesawa1997}.
Apart from the 2016/2017 outburst, no other such bright and long-lived outburst has been observed since the previous one in late 2005--2007. 
Our results are also consistent with other evidence based on polarimetry iin 2016 that the outburst originated in the central region, at a projected size of $\leq$0.67 pc  \citet{Myserlis2018}. 

In the 2010s, for the first time we achieved a very dense light curve coverage beyond the optical band, including several UV bands and X-rays up to 10 keV (in addition to multiple optical bands, and multiple radio frequencies). Therefore, predictions of a recurrence of a similar outburst in a time frame of 10--12 years from now will be testable in many wavebands.
Given the complex gas physics and magnetic fields involved, the best way to identify binaries will continue to be a phenomenological one based on semi-periodicities [or direct spatial resolution of the components, {\em{if}} both of them are radio emitting \citep[][]{Rodriguez2006, KZ2016}]. Possible emission mechanisms within the  OJ 287 binary scenarios to explain the 2016/2017 outburst
will be discussed further in future work.

In summary, we followed a phenomenological approach in identifying the latest double-peaked outburst, since  high-precision modelling is 
still challenging given the huge numbers of free parameters and the complicated
physics of this problem of a gas-rich binary system with disk impacts, and where strict periodicity
is not expected even if the secondary is not highly precessing. Additional motivation for our approach came from the fact the  
high-precision prediction of the PB model of characteristic flaring activity in 2021 
was not confirmed by our observations. 
The identification of the 2016-2017 outburst 
as the latest optical double-peaked outburst of OJ 287  is consistent with alternative binary models, matching several 
aspects of the models of, e.g.,\citet{Valtaoja2000} and  \citet{Liu2002}. 
A revision of those models would still be needed to account for the fact, that the latest  bursts 
came early, 
and for the fact that the 
whole double-peaked multi-months burst structure is associated with radio emission. Such a revised model is beyond the scope of this publication, and will be discussed in future work. 

Our new prediction is, that we should observe a new double-peaked outburst with properties matching the characteristic
X-ray--UV--optical and radio properties of the 2016-2017 outburst, and with the first peak in 2026-2028. In contrast, the PB model predicted the latest outburst to happen in October 2022 \citep{Valtonen2022}. At that epoch we found OJ 287 in a low-state instead (Komossa et al. 2023).  

Finally, it is interesting to note that the brightness of past double-peaked outbursts has systematically declined over the last few decades \citep[][]{Valtaoja2000, Dey2018, Komossa2021c}, no longer reaching the exceptional magnitudes of the 1970s and 1980s, and therefore making it more difficult to distinguish between blazar flaring in general, and binary-driven flaring in particular, based on peak flux alone.
However, including pre-1970 data in the long-term light curve, a possible period of $\sim$60 years in the brightness of double peaks has been identified \citep{Valtonen2006}. If it continues, we then expect that future double peaks will become systematically brighter again.    

\section{Summary and conclusions}
\label{sec:summary}

We have presented densely covered multifrequency radio light curves of OJ 287 between 2015 and 2022, along with optical, UV, X-ray and $\gamma$-ray observations. All data except in the $\gamma$-ray band were obtained by us in the course of the project MOMO. 
OJ 287 displayed a broad range of extraordinary activity 
including multiple flare events, deep fades, and strong spectral variability.  
Our main results can be summarized as follows: 

\begin{itemize}
\item  We have characterized in detail the radio flux and spectral variability of OJ287 between 2015 and 2022, including turn-over frequencies, spectral indices, fractional variability amplitudes and DCFs. 
In particular, we densely cover the large nonthermal 2016/2017 MWL outburst that is accompanied by strong radio flaring.

\item We find that deep low-states repeat every 1--2 years in the optical and radio regime. 
Independent of their nature, predicting the next such events is important, as optical deep fades will facilitate host galaxy imaging spectroscopy for independent SMBH mass measurements, e.g., via stellar velocity dispersion. 

\item The two brightest $\gamma$-ray flares in recent years coincide with the sharp rise and re-rise of the bright 2021--2022 radio flare, suggesting a causal connection. 

\item The light curve evolution and spectroscopic information have been used to test predictions of binary SMBH models. Precursor flare activity, predicted by the precessing binary model to occur in 2021 December, is absent. Neither the flare, nor the thermal bremsstrahlung spectrum were observed; neither in 2021 December nor any other time until 2022 June. 

\item 
We interpret the big 2016/2017 outburst as the latest of the characteristic semi-periodic double-peaked outbursts 
that OJ 287 is famous for. 
This favors binary SMBH models of OJ 287 that involve a non-(or mildly)precessing binary. 
 A scenario involving a highly precessing binary (PB model) with a particularly high primary SMBH mass of order 10$^{10}$ M$_{\odot}$ is then no longer required.

This interpretation leads to the prediction of the next double-peaked outburst in the period 2026--2028. Instead, the PB model predicted this outburst to happen in 2022 October. However, no outburst was observed at that time \citep{Komossa2023}. 

\end{itemize}

The MOMO project continues with the ultimate goal of covering one to two decades of dense multiwavelength observations of OJ 287. 

% Authors are encouraged to use an online tool at \url{http://authortools.aas.org/FIGSETS/make-figset.html} to generate their own specific figure set mark up to incorporate into their \latex\ articles.

\begin{acknowledgments} 
It is our great pleasure to thank the Swift team for carrying out the observations of OJ 287 that we proposed, and for very useful discussions regarding the observational set-up and related questions throughout the years.
This work is partly based on data obtained with the 100-m telescope of the Max-Planck-Institut f\"ur Radioastronomie at Effelsberg.
The Submillimeter Array near the summit of Maunakea is a joint project between the Smithsonian Astrophysical Observatory and the Academia Sinica Institute of Astronomy and Astrophysics and is funded by the Smithsonian Institution and the Academia Sinica. 
The authors recognize and acknowledge the very significant cultural role and reverence that the summit
of Maunakea has always had within the indigenous Hawaiian community. We are most fortunate to have the opportunity to conduct
observations from this mountain.
This research has made use of the
XRT Data Analysis Software (XRTDAS) developed under the responsibility
of the ASI Science Data Center (SSDC), Italy.     
This work has made use of Fermi-LAT data supplied by
\citet{Kocevski2021} at \url{https://fermi.gsfc.nasa.gov/ssc/data/access/lat/LightCurveRepository/}.
This research has made use of the NASA/IPAC Extragalactic Database (NED) which is operated by the Jet Propulsion Laboratory, California Institute of Technology, under contract with the National Aeronautics and Space Administration.
\end{acknowledgments}

%% To help institutions obtain information on the effectiveness of their 
%% telescopes the AAS Journals has created a group of keywords for telescope 
%% facilities.
%
%% Following the acknowledgments section, use the following syntax and the
%% \facility{} or \facilities{} macros to list the keywords of facilities used 
%% in the research for the paper.  Each keyword is check against the master 
%% list during copy editing.  Individual instruments can be provided in 
%% parentheses, after the keyword, but they are not verified.

\vspace{5mm}
\facilities{Effelsberg 100\,m radio telescope, SMA, Neil Gehrels Swift observatory (XRT and UVOT), Fermi.}

%% Similar to \facility{}, there is the optional \software command to allow 
%% authors a place to specify which programs were used during the creation of 
%% the manuscript. Authors should list each code and include either a
%% citation or url to the code inside ()s when available.

\software{HEASoft (\url{https://heasarc.gsfc.nasa.gov/docs/software/heasoft/}) with XSPEC \citep{Arnaud1996},  ESO-MIDAS (\url{https://www.eso.org/sci/software/esomidas/}), 
the R programming language (\url{https://www.r-project.org/}),
SFA \citep[][\url{https://github.com/Starkiller4011/SFA}]{Gallo2018}, 
and Python (\url{https://www.python.org/}). 
}

%% Appendix material should be preceded with a single \appendix command.
%% There should be a \section command for each appendix. Mark appendix
%% subsections with the same markup you use in the main body of the paper.

\appendix

\section{ACF} 
\label{sec:appendix-ACF}

\begin{figure}[b]
\centering
\includegraphics[clip, width=9cm]{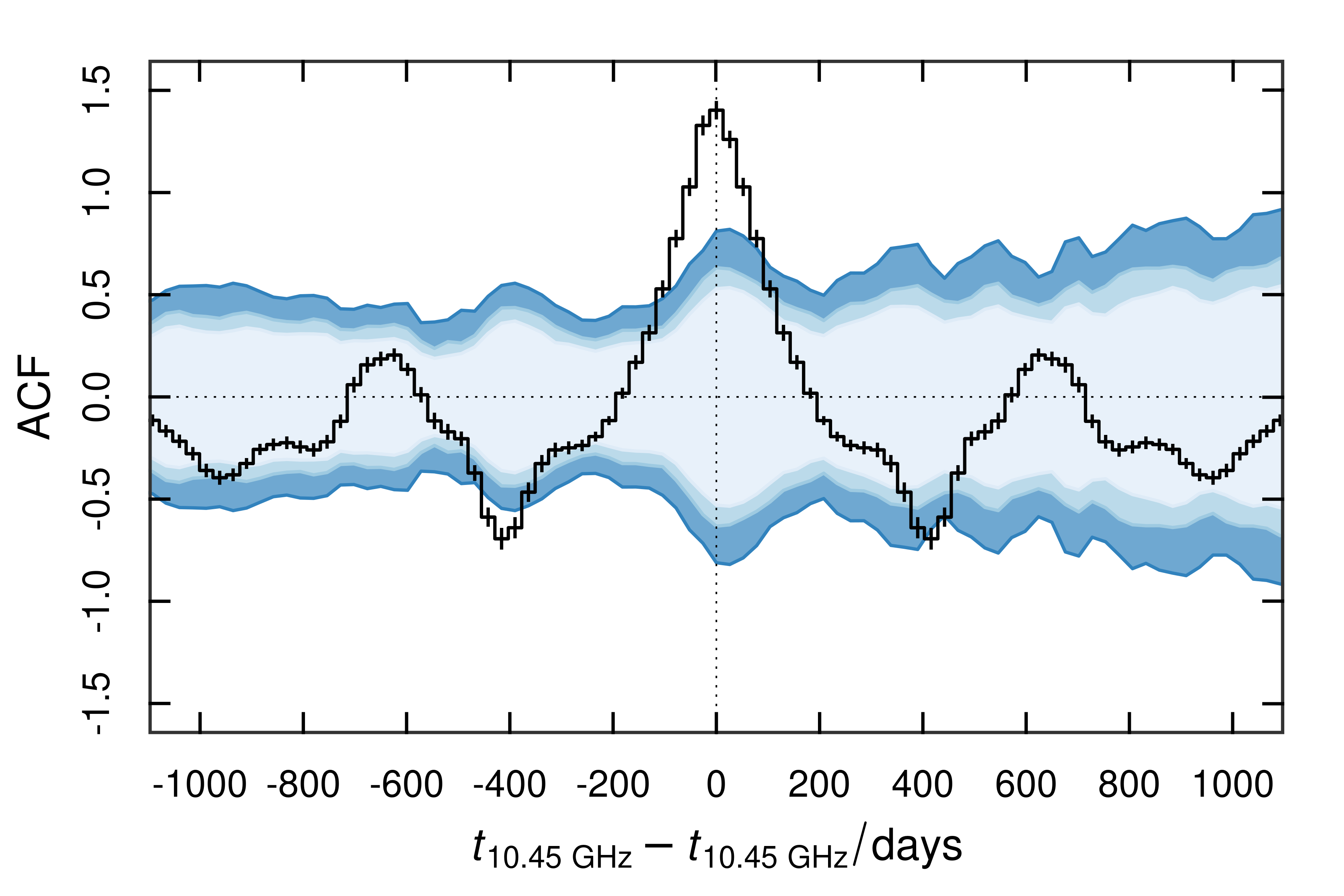}
    \caption{Long-term ACF at 10.45 GHz. }
    \label{fig:ACF}
\end{figure}

The autorcorrelation function (ACF) for the reference band of 10.45 GHz is shown in Fig. \ref{fig:ACF}. 

\section{MWL light curves}
\begin{figure*}
\centering
\includegraphics[clip, trim=0.9cm 5.3cm 1.0cm 2.3cm, width=16cm]{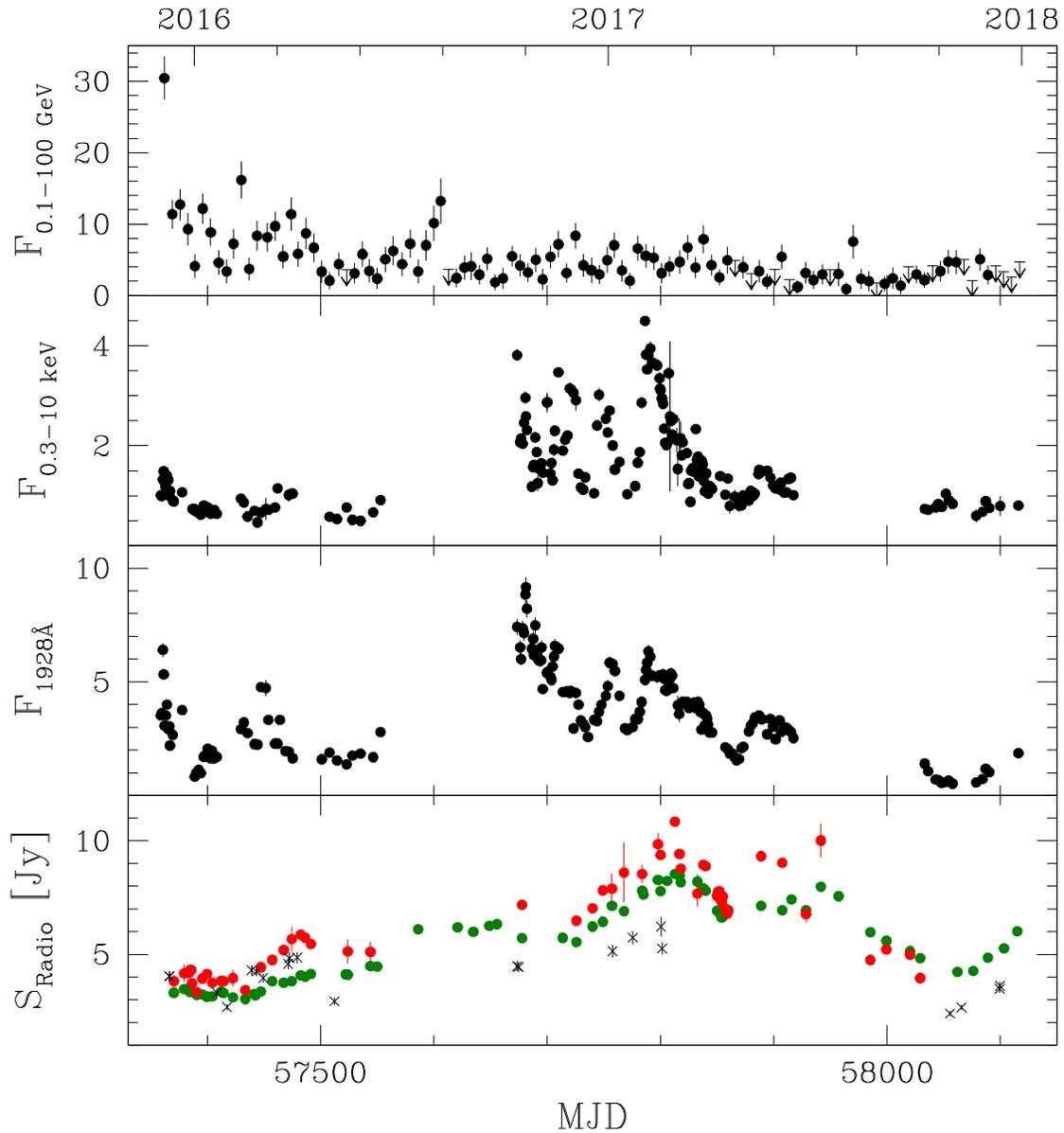}
    \caption{As Fig. \ref{fig:MWL-all} but now zoomed on the time interval 2015 December -- 2017 December, including the bright 2016/2017 radio--X-ray outburst that does not have a significant counterpart in $\gamma$-rays.  }
    \label{fig:MWL-2017}
\end{figure*}

\begin{figure*}
\centering
\includegraphics[clip, trim=0.9cm 5.3cm 1.0cm 2.3cm, width=16cm]{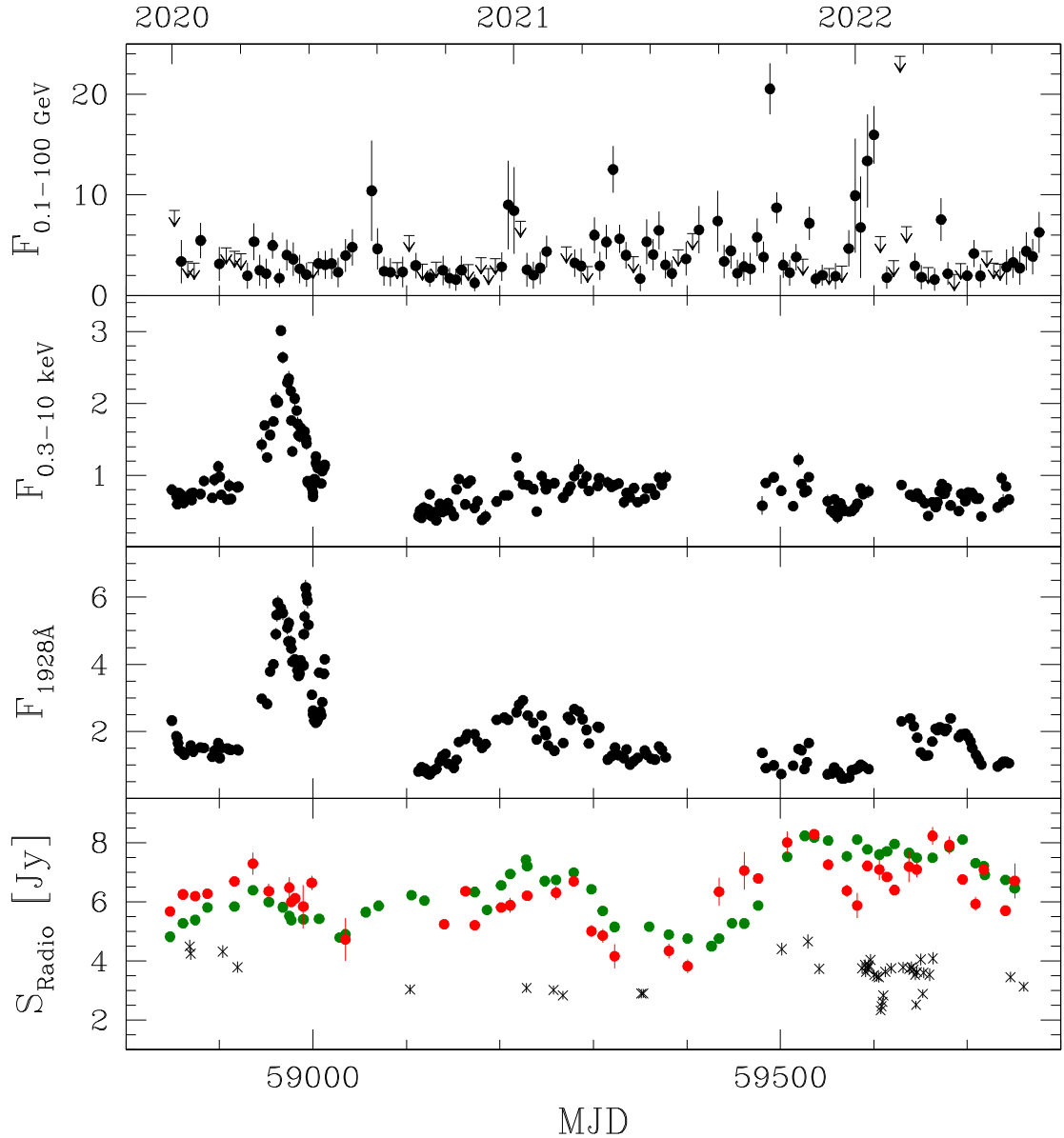}
    \caption{As Fig. \ref{fig:MWL-all} (note the change in the scale of the ordinate to match the dynamic range of the data) but now zoomed on the time interval 2020 January -- 2022 June, including the
    second-brightest radio flare peaking around 2021 November--2022 May. 
    The gap in Swift observations between 2022 mid-January to mid-February is due to a satellite safe mode during which no data were taken.
    }
    \label{fig:MWL-2022}
\end{figure*}

MWL light curves are shown here in higher resolution, zooming on selected time intervals around the main outbursts (Fig. \ref{fig:MWL-2017} and \ref{fig:MWL-2022}). 

\section{Radio data} 
Effelsberg and SMA flux density measurements are reported in Tab. \ref{tab:EB-data} and \ref{tab:SMA-data}. The full tables will be available in the online material.

\begin{table}
%\scriptsize
\centering
\caption{Radio flux density measurements between 1.4 and 44 GHz obtained at the Effelsberg telescope in the course of the MOMO project between 2017 January and 2022 June. Effelsberg data from the same project between 2015 December and 2016 December were already presented in \citet{Myserlis2018}. The columns are (1) radio frequency in GHz, (2) modified Julian date, (3) flux density in Jansky and (4) its error. The full table is available in the online material.}
\label{tab:EB-data}
\begin{tabular}{llll} 
		\hline
$\nu$ &  MJD  & $S_{\nu}$   & err \\ 
(1)   & (2)   & (3)         & (4)   \\
   	\hline
 1.4 & 58609.870 & 2.381 & 0.018\\
 1.4 & 59315.008 & 2.253 & 0.028\\
 1.4 & 59453.646 & 2.033 & 0.043\\
 1.4 & 59597.053 & 2.561 & 0.019\\
 1.4 & 59667.834 & 3.018 & 0.022\\
 1.4 & 59752.623 & 3.061 & 0.033\\    
     \hline  
 2.64 & 57767.858 & 2.789 & 0.010\\
 ... &  &  & \\
    \hline
\end{tabular}
\end{table}

\begin{table}
%\scriptsize
\centering
\caption{Radio flux density measurements from the SMA between 2015 October and 2022 June. The columns are (1) modified Julian date, (2) radio band, (3) radio frequency in GHz, (4) flux density in Jansky measured at that frequency and (5) its error. The full table is available in the online material.}
\label{tab:SMA-data}
\begin{tabular}{lllll} 
		\hline
MJD &    band  & $\nu$    & $S_{\nu}$   & err \\ 
(1) &  (2)   & (3) &   (4) &  (5) \\
   	\hline
57309.637  & 1.3 mm & 224.87 & 1.975 & 0.116 \\
57313.605  & 1.3 mm & 225.45 & 2.218 & 0.126 \\
57327.535  & 1.3 mm & 224.83 & 2.760 & 0.153 \\
57333.574  & 1.3 mm & 225.46 & 3.016 & 0.152 \\
57344.480  & 1.3 mm & 224.95 & 3.208 & 0.162 \\
... &  &  &  &  \\
		\hline
\end{tabular}
\end{table}

%\section{Author publication charges} \label{sec:pubcharge}

%\bibliography{sample631}{}

\begin{thebibliography}{99}

\bibitem[\protect\citeauthoryear{Abdo et al.}{2009}]{Abdo2009}
Abdo A.A., et al. 2009, ApJ 700, 597

\bibitem[\protect\citeauthoryear{Abdo et al.}{2010}]{Abdo2010}
Abdo A.A., et al. 2010, ApJ 716, 30

\bibitem[\protect\citeauthoryear{Abdollahi et al.}{2022}]{Abdollahi2022}
Abdollahi S., et al. 2022, ApJS 260, 53

\bibitem[\protect\citeauthoryear{Arnaud}{1996}]{Arnaud1996}
Arnaud K.A., 1996, ASPC 101, 17 
%1996ASPC..101...17A 

\bibitem[\protect\citeauthoryear{Atwood et al.}{2009}]{Atwood2009}
Atwood W.B., et al. 2009, ApJ 697, 1071 

\bibitem[\protect\citeauthoryear{Blandford et al.}{2019}]{Blandford2019}
Blandford R.D., Meier D., Readhead A., 2019, ARA\&A 57, 467 

\bibitem[\protect\citeauthoryear{Bhatta et al.}{2016}]{Bhatta2016}
Bhatta G., et al. 2016, ApJ 832, 47

\bibitem[\protect\citeauthoryear{B\"ottcher}{2019}]{Boettcher2019}
B\"ottcher M., 2019, Galaxies 7, 20

\bibitem[\protect\citeauthoryear{Britzen et al.}{2018}]{Britzen2018}
Britzen S., et al. 2018, MNRAS 478, 3199

\bibitem[\protect\citeauthoryear{Burrows et al.}{2005}]{Burrows2005} Burrows D.N., et al. 2005, SSRv 120, 165


\bibitem[\protect\citeauthoryear{Dey et al.}{2018}]{Dey2018}  Dey L., et al. 2018, ApJ 866, 11


\bibitem[\protect\citeauthoryear{Edelson \& Krolik}{1988}]{Edelson1988}
Edelson R.A., Krolik J.H., 1988, ApJ 333, 646  

\bibitem[\protect\citeauthoryear{Edelson et al.}{2002}]{Edelson2002}
Edelson R., et al. 2002, ApJ 568, 610

\bibitem[\protect\citeauthoryear{Fuhrmann et al.}{2016}]{Fuhrmann2016} Fuhrmann L. et al. 2016, A\&A 596A, 45

\bibitem[\protect\citeauthoryear{Gallo et al.}{2018}]{Gallo2018}
Gallo L.C., et al. 2018, MNRAS 478, 2557

\bibitem[\protect\citeauthoryear{Gehrels et al.}{2004}]{Gehrels2004}
Gehrels N., et al. 2004, ApJ 611, 1005

\bibitem[\protect \citeauthoryear{Ghisellini et al.}{2017}]{Ghisellini2017} Ghisellini G., Righi C., Costamante L., Tavecchio F.,
2017, MNRAS 469, 255
 
\bibitem[\protect\citeauthoryear{Goddi et al.}{2021}]{Goddi2021}
Goddi C., et al. 2021, ApJL 910, L14 

\bibitem[\protect\citeauthoryear{Gurwell et al.}{2007}]{Gurwell2007}
Gurwell M.A., Peck A.B., Hostler S.R., Darrah, M.R., Katz C.A., 2007, in From Z-Machines to ALMA: (Sub)Millimeter Spectroscopy of Galaxies, ASPC 375, 234

\bibitem[\protect\citeauthoryear{Hodgson et al.}{2017}]{Hodgson2017}
Hodgson J.A., et al. 2017, A\&A 597, A80

\bibitem[\protect\citeauthoryear{Hovatta et al.}{2008}]{Hovatta2008}
Hovatta T., Lehto H.J., Tornikoski M., 2008, A\&A 488, 897

\bibitem[\protect\citeauthoryear{Hughes et al.}{1998}]{Hughes1998} Hughes P.A., Aller H.D., Aller M.F., 1998, ApJ 503, 662

\bibitem[\protect\citeauthoryear{Idesawa et al.}{1997}]{Idesawa1997} Idesawa E., et al. 1997, PASJ 49, 631

\bibitem[\protect\citeauthoryear{Kammoun et al.}{2021}] {Kammoun2021}
Kammoun E.S., et al. 2021, MNRAS 503, 416

\bibitem[\protect\citeauthoryear{Kapanadze et al.}{2018}] {Kapanadze2018}
Kapanadze B., et al. 2018, MNRAS 480, 407

\bibitem[\protect\citeauthoryear{Katz}{1997}]{Katz1997}
Katz J.I., 1997, ApJ 478, 527

\bibitem[\protect\citeauthoryear{Kinman et al.}{1971}]{Kinman1971}
Kinman K., 1971, ApL 9, 147

\bibitem[\protect\citeauthoryear{Kocevski et al.}{2021}]{Kocevski2021}
Kocevski D., Valverde J., Garrappa S., Negro M., Brill A., Ballet J., Lott B., 2021, Astron. Telegram 15110 

\bibitem[\protect\citeauthoryear{Komossa \& Zensus}{2016}]{KZ2016}
Komossa S., Zensus J.A., 2016, IAUS 312, 13

\bibitem[\protect\citeauthoryear{Komossa et al.}{2017}]{Komossa2017}
Komossa S., et al. 2017, IAUS 324, 168 (paper Ia)

\bibitem[\protect\citeauthoryear{Komossa et al.}{2020}]{Komossa2020}
 Komossa S., et al. 2020, MNRAS 498, L35 (paper II) 

\bibitem[\protect\citeauthoryear{Komossa et al.}{2021a}]{Komossa2021a}
Komossa S., et al. 2021a, MNRAS 504, 5575 (paper III) 

\bibitem[\protect\citeauthoryear{Komossa et al.}{2021b}]{Komossa2021b}
Komossa S., et al. 2021b, Publ. Astron. Obs. Belgrade 100, 29; arXiv:2104.12901

\bibitem[\protect\citeauthoryear{Komossa et al.}{2021c}]{Komossa2021c}
Komossa S., et al. 2021c, ApJ 923, 51  (paper IV)

\bibitem[\protect\citeauthoryear{Komossa et al.}{2021d}]{Komossa2021d}
Komossa S., et al. 2021d, Universe 7, 261

\bibitem[\protect\citeauthoryear{Komossa et al.}{2022a}]{Komossa2022a}
Komossa S., et al. 2022a, MNRAS 513, 3165  (paper V) 

\bibitem[\protect\citeauthoryear{Komossa et al.}{2022b}]{Komossa2022b}
Komossa S., et al. 2022b, proceedings of the XMM-Newton 2022 Science Workshop on Black Hole Accretion, to appear in AN; 
arXiv:2207.11291 (paper Vb) 

\bibitem[\protect\citeauthoryear{Komossa et al.}{2023}]{Komossa2023} Komossa S., et al. 2023, MNRAS Letter, in press 

\bibitem[\protect\citeauthoryear{Kraus et al.}{2003}]{Kraus2003}
Kraus A., et al. 2003, A\&A 401, 161

\bibitem[\protect\citeauthoryear{Lee et al.}{2020}]{Lee2020}
Lee J.W., et al. 2020, ApJ 902, 104

\bibitem[\protect\citeauthoryear{Lehto \& Valtonen}{1996}]{Lehto1996}
Lehto H.J., Valtonen M.J., 1996, ApJ 460, 207

\bibitem[\protect\citeauthoryear{Liska et al.}{2018}]{Liska2018}
Liska M., Hesp C., Tchekhovskoy A., Ingram A., van der Klis M., Markoff S., 
2018, MNRAS, 474, L81

\bibitem[\protect\citeauthoryear{Lister et al.}{2021}]{Lister2021}
Lister M., et al. 2021, ApJ 923, 30 

\bibitem[\protect\citeauthoryear{Liu \& Wu}{2002}]{Liu2002}
Liu F.K., Wu X.B., 2002, A\&A, 388, L48

\bibitem[\protect\citeauthoryear{Lynden-Bell}{1969}]{Lynden-Bell1969}
Lynden-Bell D., 1969, Nature 223, 690 

\bibitem[\protect\citeauthoryear{Myserlis et al.}{2018}]{Myserlis2018}
Myserlis I., et al. 2018, A\&A 619, A88 (paper Ib)

\bibitem[\protect\citeauthoryear{Nilsson et al.}{2010}]{Nilsson2010}
Nilsson K., et al. 2010, A\&A 516, A60 

\bibitem[\protect\citeauthoryear{Nilsson et al.}{2020}]{Nilsson2020}
Nilsson K., et al. 2020, ApJ 904, 102

\bibitem[\protect\citeauthoryear{O'Brien}{2017}]{OBrien2017}
O'Brien S., 2017, Proc. 35th International Cosmic Ray Conference (ICRC 2017), arXiv:1708.02160

\bibitem[\protect\citeauthoryear{Ott et al.}{1994}]{Ott1994}
Ott M., et al. 1994, A\&A 284, 331

\bibitem[\protect\citeauthoryear{Peterson}{1998}]{Peterson1998}
Peterson B.M., 1998, PASP 110, 660

\bibitem[\protect\citeauthoryear{Pihajoki et al.}{2013}]{Pihajoki2013}
Pihajoki P., et al. 2013, ApJ 764, 5

\bibitem[\protect\citeauthoryear{Prince et al.}{2021}]{Prince2021}
Prince R., et al. 2021, MNRAS 508, 315 

\bibitem[\protect\citeauthoryear{Rieger}{2004}]{Rieger2004} Rieger F.M., 2004, ApJL, L5  

\bibitem[\protect\citeauthoryear{Rodriguez et al.}{2006}]{Rodriguez2006}
Rodriguez C., et al. 2006, ApJ 646, 49 

\bibitem[\protect\citeauthoryear{Roming et al.}{2005}]{Roming2005}
Roming P.W.A., et al. 2005, SSRv 120, 95

\bibitem[\protect\citeauthoryear{Sandrinelli et al.}{2016}]{Sandrinelli2016} Sandrinelli A., Covino S., Dotti M., Treves A., 2016, AJ 151, 54 

\bibitem[\protect\citeauthoryear{Schleicher et al.}{2019}]{Schleicher2019} Schleicher B., et al. 2019, Galaxies 7(2), 62

\bibitem[\protect\citeauthoryear{Sillanp\"a\"a et al.}{1988}]{Sillanpaa1988}
Sillanp\"a\"a A., et al. 1988, ApJ 325, 628

\bibitem[\protect\citeauthoryear{Sillanp\"a\"a et al.}{1996}]{Sillanpaa1996}
Sillanp\"a\"a A., et al. 1996, A\&A 315, L13

\bibitem[\protect\citeauthoryear{Singh et al.}{2022}]{Singh2022}
Singh K.P., et al. 2022, MNRAS 509, 26
% , eprint arXiv:2110.14978 

\bibitem[\protect\citeauthoryear{Sitko \& Junkkarinen}{1985}]{SitkoJunkkarinen1985}
Sitko M.L., Junkkarinen V.T., 1985, PASP 97, 1158
   
\bibitem[\protect\citeauthoryear{Timmer \& K\"onig }{1995}]{Timmer1995}
Timmer J., K\"onig M., 1995, A\&A 300, 707 

\bibitem[\protect\citeauthoryear{Valtaoja et al.}{2000}]{Valtaoja2000}
Valtaoja E., et al. 2000, ApJ 531, 744

\bibitem[\protect\citeauthoryear{Valtonen et al.}{2006}]{Valtonen2006}Valtonen M.J., et al. 2006, ApJ 646, 36

\bibitem[\protect\citeauthoryear{Valtonen et al.}{2021}]{Valtonen2021}
Valtonen M.J., et al. 2021, Galaxies, 10, 1 

\bibitem[\protect\citeauthoryear{Valtonen et al.}{2022}]{Valtonen2022}
Valtonen M.J., et al. 2022; eprint arXiv:2209.08360 (version 1)

\bibitem[\protect\citeauthoryear{Vaughan et al.}{2003}]{Vaughan2003} 
Vaughan S., Edelson R., Warwick R.S., Uttley, P., 2003, MNRAS, 345, 1271

\bibitem[\protect\citeauthoryear{Vedantham et al.}{2017}]{Vedantham2017} 
Vedantham H.K., et al. 2017, ApJ 845, 89 

\bibitem[\protect\citeauthoryear{Villata et al.}{1998}]{Villata1998}
Villata M., et al. 1998, MNRAS 293, L13

\bibitem[\protect\citeauthoryear{Villforth et al.}{2010}]{Villforth2010}
Villforth C., et al. 2010, MNRAS 402, 2087
 

\bibitem[\protect\citeauthoryear{Weaver et al.}{2022}]{Weaver2022}
Weaver Z.R., et al. 2022, ApJS 260, 12

\bibitem[\protect\citeauthoryear{Wright}{2006}]{Wright2006}
Wright E.L., 2006, PASP 118, 1711

\bibitem[\protect\citeauthoryear{Yao \& Komossa}{2021}]{YaoKomossa2021} Yao S., Komossa S., 2021, MNRAS 501, 1384

\bibitem[\protect\citeauthoryear{Zhao et al.}{2022}]{Zhao2022}
Zhao G.-Y.,  et al. 2022, ApJ 932, 72

\end{thebibliography}
%\bibliographystyle{aasjournal}

%% This command is needed to show the entire author+affiliation list when
%% the collaboration and author truncation commands are used.  It has to
%% go at the end of the manuscript.
%\allauthors

%% Include this line if you are using the \added, \replaced, \deleted
%% commands to see a summary list of all changes at the end of the article.
%\listofchanges

\end{document}